\documentclass[twocolumn]{aastex631}

\graphicspath{{./}{figures/}}

\begin{document}

\title{High-Resolution [\ion{C}{1}] \(({^3\mathrm{P}_2}\rightarrow{^3\mathrm{P}_1})\) and CO \((J=7\rightarrow6)\) Observations of Circumnuclear Disks in Nearby Seyfert Galaxies}

\correspondingauthor{Dragan Salak}
\email{dragan@astr.tohoku.ac.jp}

\author[0000-0002-3848-1757]{Dragan Salak}
\affiliation{Astronomical Institute, Tohoku University, 6-3 Aramaki, Aoba-ku, Sendai, Miyagi 980-8578, Japan}

\author[0009-0007-4497-3954]{Suphakorn Suphapolthaworn}
\affiliation{Department of Cosmosciences, Graduate School of Science, Hokkaido University, Kita 10 Nishi 8, Kita-ku, Sapporo, Hokkaido 060-0810, Japan}

\author[0000-0002-7616-7427]{Yusuke Miyamoto}
\affiliation{Department of Electrical, Electronic and Computer Engineering, Fukui University of Technology, 3-6-1 Gakuen, Fukui, Fukui 910-8505, Japan}

\author[0000-0002-5461-6359]{Naomasa Nakai}
\affiliation{School of Science and Technology, Kwansei Gakuin University, 2-1 Gakuen, Sanda, Hyogo 669-1337, Japan}

\author{Masumichi Seta}
\affiliation{School of Science and Technology, Kwansei Gakuin University, 2-1 Gakuen, Sanda, Hyogo 669-1337, Japan}

\author[0000-0003-1420-4293]{Kazuo Sorai}
\affiliation{Department of Cosmosciences, Graduate School of Science, Hokkaido University, Kita 10 Nishi 8, Kita-ku, Sapporo, Hokkaido 060-0810, Japan}
\affiliation{Department of Physics, Faculty of Science, Hokkaido University, Kita 10 Nishi 8, Kita-ku, Sapporo, Hokkaido 060-0810, Japan}

\begin{abstract}

We present [\ion{C}{1}] \(({^3\mathrm{P}_2}\mathrm{-}{^3\mathrm{P}_1})\), CO \((J=7\mathrm{-}6)\), and 800-GHz continuum observations of the circumnuclear disks (CNDs) in the Seyfert galaxies NGC 613 and NGC 1808 with the Atacama Large Millimeter/submillimeter Array. The images reveal the distributions of neutral gas and dust in nearby galaxy nuclei at a resolution of \(0\farcs2\) (10 pc) which is unprecedented at these frequencies. Both lines and continuum exhibit peaks at the position of molecular tori. The maximum [\ion{C}{1}]\(({^3\mathrm{P}_2}\mathrm{-}{^3\mathrm{P}_1})/({^3\mathrm{P}_1}\mathrm{-}{^3\mathrm{P}_0})\) integrated-intensity ratio is \(\approx1.1\) (\(\mathrm{K~km~s}^{-1}\) scale) in the region of the torus, yielding an excitation temperature of \(T_\mathrm{ex}\approx60~\mathrm{K}\) under the approximation of local thermodynamic equilibrium (LTE) and optically thin emission. A non-LTE radiative transfer analysis of the [\ion{C}{1}] line ratio in the CND constrains the density and kinetic temperature of molecular gas to \(n_\mathrm{H_2}>10^3~\mathrm{cm}^{-3}\) and \(T_\mathrm{k}>50~\mathrm{K}\), respectively. To estimate the C/CO abundance ratio and physical conditions, we simultaneously modeled the [\ion{C}{1}] and multiple CO lines, and obtained solutions with densities and temperatures of \(n_\mathrm{H_2}\sim10^{3\mathrm{-}4}~\mathrm{cm}^{-3}\) and \(T_\mathrm{k}\sim80\mathrm{-}250~\mathrm{K}\) in the central 63 pc. The C/CO ratios are \(\approx2\) in NGC 613 and \(\approx0.6\) in NGC 1808. C/CO tends to increase with \(T_\mathrm{k}\) and decrease with \(n_\mathrm{H_2}\) in the CND, which is consistent with models of photodissociation regions. Based on the kinematics of gas traced by [\ion{C}{1}] \(({^3\mathrm{P}_2}\mathrm{-}{^3\mathrm{P}_1})\) and CO \((J=7\mathrm{-}6)\), the mass of the supermassive black hole at the center of NGC 613 is estimated to be \(\sim2\times10^7~M_\sun\).

\end{abstract}



\section{Introduction}\label{sec:intro}

The interstellar medium (ISM) in many active galactic nuclei (AGN) contains multi-phase gas concentrated in a molecular torus (size \(\sim1\mathrm{-}10~\mathrm{pc}\)) within a larger circumnuclear disk (CND; \(\sim100~\mathrm{pc}\)). The gas is feeding the supermassive black hole via inflows, whereas feedback from radiation and jets drives outflows (e.g., \citealt{Wad12,WBU12,Com19,Ima20,GB21}). To understand the processes that regulate nuclear activity, it is important to study the physical conditions and dynamics of all ISM phases in the CND, and especially neutral gas that dominates the mass budget.

Atomic carbon (C) is one of key constituents of neutral gas as it regulates cooling via the fine-structure lines and plays a major role in the chemistry involving CO. The abundance and distribution of C are therefore of importance for our understanding of the conditions in the ISM. The C/CO abundance ratio is controlled by various mechanisms (e.g., \citealt{WVC22}): photodissociation of CO by far-ultraviolet (FUV) radiation from massive stars in photodissociation regions (PDRs) where a multi-zone C\(^{+}\)/C/CO structure forms \citep{TH85,vDB88,HTT91,Kau99}, ionization by X-rays in X-ray dominated regions (XDRs) \citep{MHT96,MS05,MSI06,MSI07,HTH13,TP24}, ionization by cosmic rays \citep{Sch93,PTV04,Mei11,BPV15,BTT21,GOB19}, chemistry in metal-poor environments with reduced shielding \citep{GC16,PBZ18,HSD21,Bis25}, time-dependent chemistry in evolving clouds (e.g., \citealt{Suz92}), and mechanical heating.

The fine structure of C is a three-level system with observable transitions \(^{3}\mathrm{P}_{1}-^{3}\mathrm{P}_{0}\) and \(^{3}\mathrm{P}_{2}-^{3}\mathrm{P}_{1}\), hereafter [\ion{C}{1}] \((1\mathrm{-}0)\) and [\ion{C}{1}] \((2\mathrm{-}1)\), respectively. Since collisions cannot excite any other level in C under typical conditions in the cold ISM, the intensities of the two lines provide a robust probe of gas conditions. The ionization energy of C is 11.3 eV, less than that of hydrogen (13.6 eV), but similar to the dissociation energy of CO in the ground state (11.1 eV). Moreover, the critical densities of [\ion{C}{1}] \((1\mathrm{-}0)\) are \(\approx490~\mathrm{cm}^{-3}\) for collisions with H atoms, and  \(\approx1000~\mathrm{cm}^{-3}\) for collisions with H\(_2\) molecules at \(T_\mathrm{k}=50~\mathrm{K}\).\footnote{The atomic and molecular data in this paper are taken from the Leiden Atomic and Molecular Database \citep{Sch05}.} By comparison, the critical density of [\ion{C}{1}] \((2\mathrm{-}1)\) under the same conditions is \(\approx3000~\mathrm{cm}^{-3}\). This is comparable to those of the CO transitions \(J=1\mathrm{-}0\) and \(J=2\mathrm{-}1\), suggesting that [\ion{C}{1}] and low-\(J\) CO lines may trace the same gas phase.

[\ion{C}{1}] observations of Galactic clouds have revealed that the distribution of C largely coincides with that of CO even inside molecular clouds where the C/CO abundance ratio is \(\sim0.02-0.4\) with an average of \(\sim0.1\) \citep{Phi80,PH81,KBP85,ZBG86,Gen88,Fre89,WP91,Her92,Min94,Plu99,Plu00,Tau95,Ike99,Mae99,Tat99,How00,Ohj01,Oka01,Shi13,Beu14}. Higher C/CO of \(\sim0.3\mathrm{-}0.5\) in the Galactic center region \citep{Tan11}, some PDRs \citep{PJK94}, and \(\gtrsim1\) at cloud edges \citep{Ike02,Bur15,OO25}, in translucent clouds \citep{SvD94,Oka05}, and in the Galactic CND \citep{TNK21} have been reported. Some molecular clouds, that have measurements of both lines, exhibit high excitation temperatures of \(30\mathrm{-}100~\mathrm{K}\), indicating heating in star-forming regions \citep{Jaf85,ZBG86,WS95,Yam01,PB15,Lee22}. Most observations are generally in accordance with clumpy PDR models \citep{MT93,Spa96}, and the mixing of C and CO inside molecular clouds may be a result of turbulence \citep{Off14,Glo15}.

Following the first detection in an external galaxy \citep{But92}, observations at low and moderate spatial resolution have revealed bright [\ion{C}{1}] emission in the central regions of many nearby galaxies \citep{Sch93,Whi94,Har95,Stu97,IB01,IB02,IB03,PG04,IWB95,Hit08,Isr14,IRW15,Isr17,Cro19,Miy21,MA23}. A notable example is M82, where early studies have found evidence of a high (\(\approx0.5\)) C/CO abundance ratio, possibly a result of a starburst \citep{Sch93,Whi94,Stu97}. Resolved observations have revealed that [\ion{C}{1}] lines exhibit a near-linear correlation with CO \((1\mathrm{-}0)\) at kpc scales \citep{Jia17,Jia19,JGZ21,MA23}. The [\ion{C}{1}] lines have been considered as probes of molecular gas mass, especially at high redshift, where standard tracers, such as CO \((1\mathrm{-}0)\), are difficult to observe \citep{Wei03,Wei05,Wal11,AZ13,Bot17,Pop17,Emo18,Ote18,Dan19,Nes19,HW20,Mad20,Val18,Val20,Lee21,Dun22,Ber23,Gur23,Li24,FC25,Top25}.

\begin{figure*}[ht!]
\epsscale{1}
\plotone{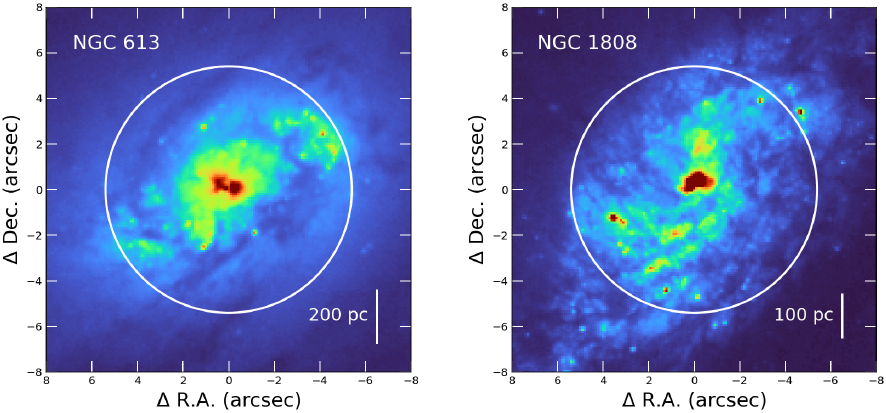}
\caption{Central regions of the target galaxies imaged by the Hubble Space Telescope (F814W filter; Hubble Legacy Archive). The field of view in observations with ALMA is shown as a white circle.}\label{fig:gals}
\end{figure*}

\begin{deluxetable*}{lccc}[ht!]
\tablecaption{Galaxy Parameters\label{tab:tar}}
\tablewidth{0pt}
\tablehead{
\colhead{} & \colhead{NGC 613} & \colhead{NGC 1808} & \colhead{References}
}
\startdata
Morphology & SB(rs)bc & (R)SAB(s)a & (1) \\
\(\alpha_\mathrm{ICRS}\) & \(01^\mathrm{h}34^\mathrm{m}18\fs189\) & \(05^\mathrm{h}07^\mathrm{m}42\fs329\) & (2) \\
\(\delta_\mathrm{ICRS}\) & \(-29\arcdeg25\arcmin06\farcs59\) & \(-37\arcdeg30\arcmin45\farcs85\) & (2)  \\
Activity & Sy, \ion{H}{2} &  Sy 2, \ion{H}{2} & (3,4) \\
\(\log{(L_\mathrm{X}/\mathrm{erg~s^{-1}})}\) \((2\mathrm{-}10~\mathrm{keV})\) & 41.3 & 40.2 & (5,6) \\
Position angle (degree) & \(298\) & \(324\) & (7,8) \\
Inclination (degree) & \(46\) & \(57\) & (7,8) \\
Systemic velocity (\(\mathrm{km~s}^{-1}\)) & 1471 & 998 & (7,8)  \\
Distance (Mpc) & 17.5 & 10.8 & (9)  \\
\enddata
\tablecomments{(1) \citet{deV91}, (2) \citet{Com19}, (3) NASA/IPAC Extragalactic Database, (4) \citet{VV86}, (5) \citet{Cas13}, (6) \citet{JB05}, (7) \citet{Miy17}, (8) \citet{Sal16}, (9) \citet{Tul88}.}
\end{deluxetable*}

High-resolution observations with the Atacama Large Millimeter/submillimeter Array (ALMA) have revealed strong and widespread [\ion{C}{1}] \((1\mathrm{-}0)\) emission in the central regions of the starburst galaxy NGC 253 \citep{Kri16} and Seyfert galaxies including the Circinus \citep{Izu18,Izu23}, NGC 6240 \citep{Cic18}, NGC 613 \citep{Miy18}, NGC 1808 \citep{Sal19}, NGC 7469 \citep{Izu20}, NGC 1068 \citep{Sai22}, NGC 3627 \citep{Liu23}, and NGC 5506 \citep{Tak26}. [\ion{C}{1}] \((1\mathrm{-}0)\) has also been imaged in a number of ultraluminous infrared galaxies (ULIRGs) \citep{Sai20,Mic21} including Arp 220 \citep{Ued22}. \citet{Sai22} found evidence of a high [\ion{C}{1}]\((1\mathrm{-}0)/\mathrm{CO}(1\mathrm{-}0)\) line intensity ratio in the molecular outflow in NGC 1068, indicating that the C abundance may be enhanced in AGN-driven outflows.

\begin{deluxetable*}{lcccc}[ht!]
\tablecaption{Observation Summary\label{tab:obs1}}
\tablewidth{0pt}
\tablehead{
\colhead{} & \colhead{NGC 613 (12 m)} & \colhead{NGC 613 (7 m)} & \colhead{NGC 1808 (12 m)} & \colhead{NGC 1808 (7 m)}
}
\startdata
Observing date & 2024 Sep. 16 & 2024 Jun. 19 & 2024 Sep. 5 & 2024 Jun. 19, Jul. 1 \\
On-source time (min.) & 15 & 29 & 14 & 33+34 \\
Number of antennas & 43 & 11 & 43 & 6, 10 \\
Flux \& bandpass calibrator & J2253+1608 & J2232+1143 & J0423-0120 & J0538-4405 \\
Phase calibrator & J0143-3200 & J0143-3200 & J0522-3627 & J0440-4333 \\
\enddata
\end{deluxetable*}

The bright [\ion{C}{1}] \((1\mathrm{-}0)\) and low-\(J\) CO line emission toward the CNDs in Seyfert galaxies is in line with expectations from both PDR and XDR models. However, as previous resolved observations have targeted only the ground-state transition, [\ion{C}{1}] \((1\mathrm{-}0)\), and most low-\(J\) CO lines are optically thick, the C/CO abundance ratio remains unconstrained observationally. Measuring both [\ion{C}{1}] lines would allow us to directly probe the column density of C and compare the line intensity ratios with predictions from PDR and XDR models.

In this paper, we report the first 10-pc-resolution observations of [\ion{C}{1}] \((2\mathrm{-}1)\), CO \((7\mathrm{-}6)\), and the underlying 800-GHz continuum toward the CNDs of two nearby Seyfert galaxies NGC 613 and NGC 1808 that harbor low-luminosity AGN (Figure \ref{fig:gals}, Table \ref{tab:tar}). The main goals of this study are: 1) to probe the C/CO abundance ratio and gas conditions in the galaxy nuclei, and 2) to study the gas dynamics in the CND and torus. The targets were selected because they harbor gas-rich CNDs in the central 100 pc and because ALMA data of [\ion{C}{1}] \((1\mathrm{-}0)\) and multiple low-\(J\) CO lines at comparable resolution exist allowing us to probe gas excitation and compare the line intensity ratios with model predictions. While NGC 613 exhibits radio jets and suppressed star formation in the CND \citep{Miy17}, NGC 1808 harbors a composite starburst-AGN nucleus dominated by star formation \citep{Sal17}. Thus, the two nuclei serve as testbeds for different environments.

The paper is organized as follows. In Section \ref{sec:obs}, we describe the observations and reduction of the new and auxiliary data used in the analysis. The main results, including continuum and line images and measured fluxes are given in Section \ref{sec:res}. In Section \ref{sec:gpc}, we derive the gas conditions and abundances of C and CO in the CNDs of the two galaxies using LTE and non-LTE modeling, and discuss on the possible mechanisms that control them. Finally, in Section \ref{sec:dyn}, we analyze the gas dynamics. A summary of key findings is given in Section \ref{sec:sum}.

\section{Observations and data reduction}\label{sec:obs}

\subsection{Band 10 Observations}

The ALMA Band 10 observations were carried out during cycle 10 in 2024 using 12-m and 7-m arrays in single-pointing mode (Table \ref{tab:obs1}). The 12-m baselines ranged from 15 m to 650 m for NGC 1808 and from 15 m to 484 m for NGC 613. The shortest baselines of the 7-m array were 9 m, corresponding to a maximum recoverable scale of \(\theta_\mathrm{MRS}\approx7\arcsec\).  The double-sideband receivers were tuned to simultaneously image [\ion{C}{1}] \((2\mathrm{-}1)\) at \(\nu_\mathrm{rest}=809.341970~\mathrm{GHz}\) and CO \((7\mathrm{-}6)\) at \(\nu_\mathrm{rest}=806.651806~\mathrm{GHz}\) in the upper sideband, and 800-GHz continuum. The spectral windows were centered at the rest frequencies of 804.777, 806.652, 807.997, and 809.342 GHz, each with a total bandwidth of 1.875 GHz that consists of 240 channels of width 7.8125 MHz.

The data were calibrated using the Common Astronomy Software Applications, version 6.5 (CASA; \citealt{TCT22}), with a standard pipeline. The 12-m array data were self-calibrated as part of the pipeline processing owing to the presence of luminous continuum sources in both targets. For line data, the continuum was subtracted in the \(uv\)-plane. Images were then produced by joint deconvolution and \textsf{CLEAN} procedure of 12-m and 7-m array data using the CASA task \textsf{tclean}. \textsf{CLEAN} was done separately for continuum in multi-frequency synthesis (\textsf{mfs}) mode using line-free channels, and for continuum-subtracted line data in \textsf{cube} mode. The total bandwidth used in continuum imaging after flagging the lines was 10.9 GHz in NGC 613 and 11.5 GHz in NGC 1808. Both sidebands were used for continuum imaging. To achieve optimal angular resolution and sensitivity, the weighting was set to Briggs with the robust parameter equal to 0.5, velocity resolution to \(5~\mathrm{km~s}^{-1}\) (for spectral lines), and the \textsf{CLEAN} threshold to \(2\sigma\), where \(\sigma\) was estimated from emission-free channels in first-generation images. We employed the auto-masking algorithm \textsf{auto-multithreshold} with standard parameters \citep{Kep20}. 

The resulting angular resolution and sensitivity of final images are summarized in Table \ref{tab:obs2}. The achieved angular resolution of the continuum image was \(0\farcs289\times0\farcs169\) for NGC 613 and \(0\farcs194\times0\farcs160\) for NGC 1808, corresponding to \(\approx14~\mathrm{pc}\) and \(\approx8~\mathrm{pc}\), respectively. All images presented below and used in the analysis were corrected for the primary beam attenuation. The velocities are given in radio definition with respect to the local standard of rest (LSR).

\begin{table}[ht!]
\begin{center}
\caption{Image Parameters\label{tab:obs2}}
\begin{tabular}{lcc}
\tableline\tableline
 & NGC 613 & NGC 1808 \\
\tableline
[\ion{C}{1}] \((2\mathrm{-}1)\) \\
FWHM & \(0\farcs294\times0\farcs179\) & \(0\farcs204\times0\farcs170\)  \\
P.A. (\(\arcdeg\)) & 67.0 & -80.8 \\
\(\sigma\) (\(\mathrm{mJy~beam^{-1}}\)) & 24 & 20 \\
\tableline
CO \((7\mathrm{-}6)\) \\
FWHM & \(0\farcs298\times0\farcs180\) & \(0\farcs201\times0\farcs0.196\) \\
P.A. (\(\arcdeg\)) & 65.8 & -69.3 \\
\(\sigma\) (\(\mathrm{mJy~beam^{-1}}\)) & 26 & 21 \\
\tableline
800-GHz continuum \\
FWHM & \(0\farcs289\times0\farcs169\) & \(0\farcs194\times0\farcs160\) \\
P.A. (\(\arcdeg\)) & 66.0 & -83.7 \\
\(\sigma\) (\(\mathrm{mJy~beam^{-1}}\)) & 1.0 & 1.0 \\
\tableline
\end{tabular}
\end{center}
\tablecomments{FWHM is the full-width half maximum, P.A. is the position angle of the beam, and \(\sigma\) denotes image noise in emission-free channels of \(5~\mathrm{km~s}^{-1}\) for data cubes.}
\end{table}

\subsection{Auxiliary Data}\label{subsec:arc}

In order to model the gas physical conditions using line intensity ratios (Section \ref{sec:gpc}), we collected the following archival ALMA data that have comparable angular resolution and sensitivity: [\ion{C}{1}] \((1\mathrm{-}0)\), CO \((1\mathrm{-}0)\), CO \((2\mathrm{-}1)\), and CO \((3\mathrm{-}2)\) for NGC 613, and [\ion{C}{1}] \((1\mathrm{-}0)\), CO \((2\mathrm{-}1)\), and CO \((3\mathrm{-}2)\) for NGC 1808. We acquired the data from the ALMA Archive in calibrated form and then generated images using \textsf{tclean} in CASA. Only 12-m array data were used, because the shortest baselines in those data were comparable to those of the new 12m + 7m Band 10 data.

We adjusted the \(uv\)-coverage of archival ALMA data so that all data used in the analysis have a common maximum recoverable scale, \(\theta_\mathrm{MRS}\sim \lambda/D\), where \(\lambda\) is the wavelength, and \(D\) is the shortest baseline. A threshold was set to the visibility data as \(uv\) values that match the shortest baselines of the 7-m array Band 10 data (\(>23~k\lambda\) for NGC 613 and \(>21~k\lambda\) for NGC 1808). The flagging of visibilities was done using \textsf{uvrange} in \textsf{tclean}. Since the \(uv\) distributions of each data set are different, the resulting beam sizes and shapes differ too. We smoothed all data in the image plane to a common circular beam that corresponds to the image with lowest resolution. In addition, all auxiliary data were processed using a common strategy: velocity resolution of \(5~\mathrm{km~s^{-1}}\), Briggs weighting with the robust parameter set to 0.5, and a \textsf{CLEAN} threshold of \(2\sigma\), as estimated from emission-free channels in first-generation cubes. We employed the non-interactive algorithm \textsf{auto-multithreshold} in \textsf{tclean} with standard parameters for the 12-m array. The resulting data cubes were corrected for primary beam attenuation and re-gridded to a common pixel size.

\begin{figure*}[ht!]
\epsscale{1}
\plotone{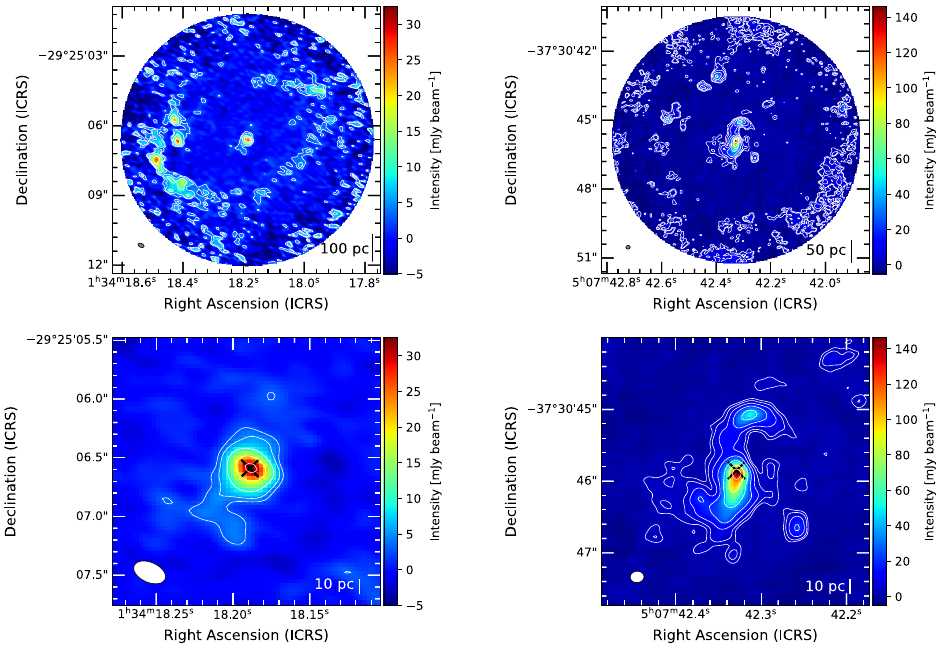}
\caption{Continuum (800 GHz) images of NGC 613 (left) and NGC 1808 (right). The contours are \((3,5,10,20,30)\sigma\) for NGC 613 and \((3,5,10,20,40,80,120)\sigma\) for NGC 1808, where \(\sigma=1~\mathrm{mJy~beam^{-1}}\). The color scale ranges from \(-5\sigma\) to the maximum value. The galaxy nucleus is indicated by an \textsf{x}-symbol. The synthesized beam is shown in the bottom left corner of each panel.}\label{fig:cont}
\end{figure*}

\break

\section{Results}\label{sec:res}

\begin{figure*}[ht!]
\epsscale{1}
\plotone{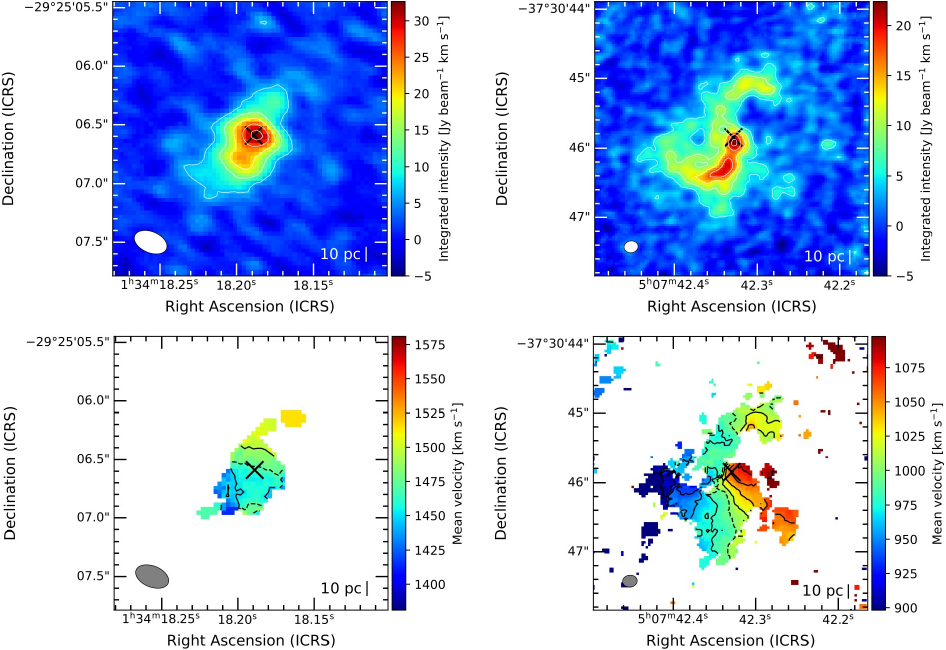}
\caption{Continuum-subtracted [\ion{C}{1}] \((2\mathrm{-}1)\) integrated intensity in NGC 613 (top left) and NGC 1808 (top right) generated with a threshold of \(-3\sigma\). The contours are plotted at \((0.2,0.4,0.6,0.8,0.95)\mathcal{I}_\mathrm{[CI](2-1)}^\mathrm{max}\) where \(\mathcal{I}_\mathrm{[CI](2-1)}^\mathrm{max}=32.6~\mathrm{Jy~beam^{-1}~km~s^{-1}}\) in NGC 613 (rms \(1.9~\mathrm{Jy~beam^{-1}~km~s^{-1}}\)) and \(22.4~\mathrm{Jy~beam^{-1}~km~s^{-1}}\) in NGC 1808 (rms \(1.5~\mathrm{Jy~beam^{-1}~km~s^{-1}}\)). The mean-velocity images are masked below a threshold of \(5\sigma\) and have contours at 1451, 1471, 1491 km s\(^{-1}\) for NGC 613 (bottom left), and from 898 to 1098 km s\(^{-1}\) in steps of 20 km s\(^{-1}\) for NGC 1808 (bottom right). The dashed contour is the adopted systemic velocity, and the galaxy center is indicated by an \textsf{x}-symbol.}\label{fig:ci21}
\end{figure*}

\subsection{Continuum Images}\label{subsec:con}

The 800-GHz continuum images of NGC 613 and NGC 1808 are shown in Figure \ref{fig:cont}. Both galaxies exhibit strong continuum emission concentrated in a point-like nucleus and extended structure in the SE-NW direction that traces the distribution of dust. The intensity distribution in NGC 1808 exhibits an \emph{S}-shaped spiral with a prominent peak at the center. The spiral appears to be trailing, as its orientation is consistent with spiral arms on kpc scale \citep{Sal16}. As shown below, a similar structure is also observed in the distribution of neutral gas.

In order to determine the position and peak intensity of the continuum emission, we performed two-dimensional fitting with a Gaussian function using the CASA task \textsf{imfit}. The center of the circular fitting region was the brightest pixel, and its radius was the minor axis (FWHM) of the beam. The derived continuum position in NGC 613 is \((\alpha,\delta)_\mathrm{ICRS}=(01^\mathrm{h}34^\mathrm{m}18\fs1887\pm0\fs0003,-29\arcdeg25\arcmin06\farcs593\pm0\farcs003)\), and the peak intensity is \(32.2\pm1.0~\mathrm{mJy~beam^{-1}}\). The derived continuum position in NGC 1808 is \((\alpha,\delta)_\mathrm{ICRS}=(05^\mathrm{h}07^\mathrm{m}42\fs3291\pm0\fs0003,-37\arcdeg30\arcmin45\farcs912\pm0\farcs005)\), and the peak intensity is \(144.3\pm5.9~\mathrm{mJy~beam^{-1}}\). The positions are offset by \(0\farcs005\) in NGC 613 and by \(0\farcs061\) in NGC 1808 relative to the coordinates of the galaxy nucleus (Table \ref{tab:tar}), which were determined from observations of the continuum at 350 GHz with a positional accuracy of \(0\farcs1\) \citep{Com19,Com26}. The peak of the 800-GHz continuum is consistent with the coordinates of an H\(_2\)O maser in NGC 613 \citep{Kon06}.

In addition to the CND, the 800-GHz continuum is also detected toward weaker sources at larger radii that are likely dusty molecular clouds associated with star formation. In NGC 613, emission is detected toward molecular clouds in the starburst ring at a radius of \(300~\mathrm{pc}\) from the center (Figure \ref{fig:cont}). The ring has been revealed and studied in detail in previous observations of molecular gas \citep{FB14,Miy17,Miy18,Aud19,Sat21,Kan23,Cho24}. Similarly, emission is detected in dusty molecular clouds at a distance of \(150~\mathrm{pc}\) from the center of NGC 1808. These clouds too can be seen in the images of [\ion{C}{1}] \((1\mathrm{-}0)\), CO lines, and millimeter and submillimeter continuum presented in previous works \citep{Sal17,Sal18,Sal19,Aud21}.

\begin{figure*}[ht!]
\epsscale{1}
\plotone{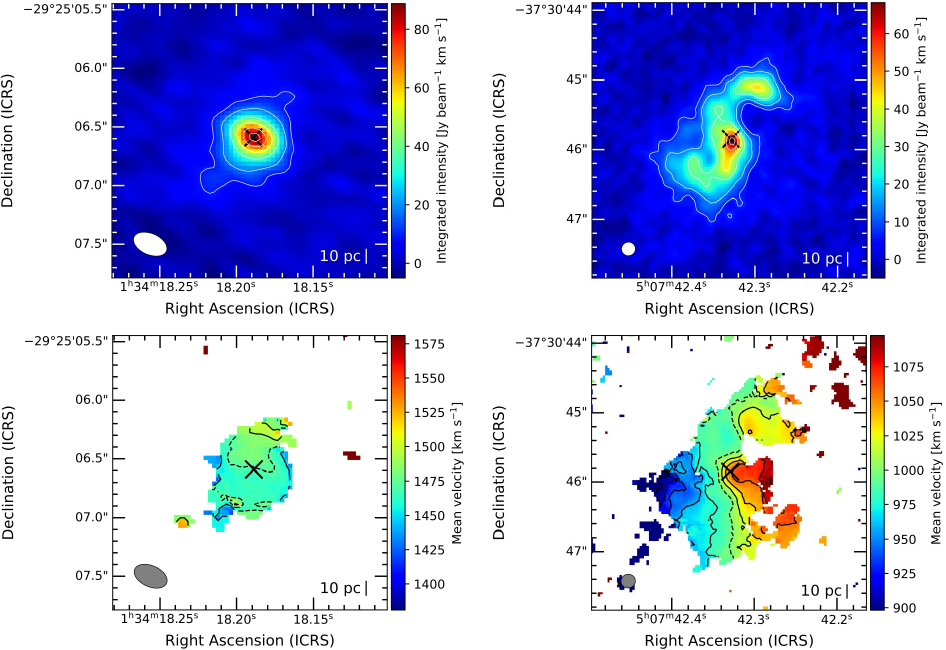}
\caption{Continuum-subtracted CO \((7\mathrm{-}6)\) integrated intensity in NGC 613 (top left) and NGC 1808 (top right) generated with a threshold of \(-3\sigma\). The contours are plotted at \((0.1,0.2,0.4,0.6,0.8,0.95)\mathcal{I}_\mathrm{CO(7-6)}^\mathrm{max}\) where \(\mathcal{I}_\mathrm{CO(7-6)}^\mathrm{max}=88.8~\mathrm{Jy~beam^{-1}~km~s^{-1}}\) in NGC 613 (rms \(2.0~\mathrm{Jy~beam^{-1}~km~s^{-1}}\)) and \(68.3~\mathrm{Jy~beam^{-1}~km~s^{-1}}\) in NGC 1808 (rms \(1.7~\mathrm{Jy~beam^{-1}~km~s^{-1}}\)). The mean-velocity images are masked below a threshold of \(5\sigma\) and have contours at 1451, 1471, 1491 km s\(^{-1}\) for NGC 613 (bottom left), and from 898 to 1098 km s\(^{-1}\) in steps of 20 km s\(^{-1}\) for NGC 1808 (bottom right). The dashed contour is the adopted systemic velocity, and the galaxy center is indicated by an \textsf{x}-symbol.}\label{fig:co76}
\end{figure*}

\subsection{[\ion{C}{1}] (2--1) Images}\label{subsec:ci21}

The distribution of [\ion{C}{1}] \((2\mathrm{-}1)\) integrated intensity (moment zero) is shown in the top panels of Figure \ref{fig:ci21}. Moment zero is defined as  \(\sum_i I_{\nu,i}\Delta v\), where \(I_\nu\) is the intensity in Jy beam\(^{-1}\) and \(v\) is the velocity. The images reveal that the atomic carbon is distributed throughout the CNDs and the peak intensity is at the position of the torus in both galaxies. The radii of molecular tori, measured from CO \((3\mathrm{-}2)\) observations, are \(8~\mathrm{pc}\) in NGC 613 and \(6~\mathrm{pc}\) in NGC 1808 \citep{Com19,Com26}, which is comparable to the ALMA beam size in our observations. The bottom panels of Figure \ref{fig:ci21} show the intensity-weighted mean velocity (moment one) defined as \(\langle{v}\rangle\equiv\sum_i I_{\nu,i}v_i/\sum_i I_{\nu,i}\). The \emph{S}-shaped spiral pattern is clearly seen in both moment zero and one images of NGC 1808, but the velocity map in the bottom left panel shows that there is a spiral pattern with the same orientation in NGC 613 too.

The velocity maps of both galaxies reveal complex gas motions in the CND. This is caused by noncircular motions, i.e., inflows and outflows. While NGC 613 exhibits outflows detected in [\ion{C}{1}] \((1\mathrm{-}0)\) and CO \((3\mathrm{-}2)\) \citep{Miy18,Aud19} in the CND, NGC 1808 exhibits inflows in the CND and outflows on larger scale \citep{Sal16,Sal17,Aud21}. Based on high-resolution molecular data, \citet{Com26} determined the position angle (PA) of the torus in NGC 613 to be \(\mathrm{PA}\approx340\arcdeg\). The [\ion{C}{1}] \((2\mathrm{-}1)\) velocity field is consistent with this picture: the kinematic PA, i.e., the normal to the dashed contour in Figure \ref{fig:ci21} is \(\approx347\arcdeg\), which is kinematically decoupled from the galaxy disk (\(298\arcdeg\); Table \ref{tab:tar}). Similarly, the kinematic position angle in the central region of NGC 1808 traced by [\ion{C}{1}]  is misaligned with the galaxy disk.

The nucleus in NGC 1808 is more than \(50~\mathrm{km~s}^{-1}\) offset from the systemic velocity of \(998~\mathrm{km~s}^{-1}\), which was derived for the CND from CO \((1\mathrm{-}0)\) observations at a resolution of \(2\arcsec\) \citep{Sal16}. This may due to the following reasons: the peak emission is not at the dynamical center (i.e., not a torus surrounding the SMBH), the SMBH itself is not at the dynamical center of the galaxy, or the kinematics is strongly affected by inflows and outflows. Since the galaxy exhibits large-scale outflows seen in \ion{Na}{1} D and CO \citep{Phi93,Sal16}, the third possibility seems plausible.

\subsection{CO (7--6) Images}\label{subsec:co76}

The images of CO \((7\mathrm{-}6)\) integrated intensity and velocity field are shown in Figure \ref{fig:co76}. In both galaxies, the intensity distribution is highly concentrated at the nucleus, but the CND in NGC 1808 exhibits extended \emph{S}-shaped distribution. The intensity peak is spatially consistent with the position of the molecular torus. It is also clear that the emission is more compact than that of [\ion{C}{1}] and concentrated in the torus in both galaxies. This is expected because the critical density of CO \((7\mathrm{-}6)\) for collisions with H\(_2\) at \(T_\mathrm{k}=50~\mathrm{K}\) is \(\approx5\times10^5~\mathrm{cm}^{-3}\), making it a tracer of very dense molecular gas.

The velocity fields (lower panels in Figure \ref{fig:co76}) are similar to that of [\ion{C}{1}]. There is evidence of both rotation (velocity gradient) and noncircular motions. The kinematic PA in the central part of the CND in NGC 613 is \(\approx350\arcdeg\), which is consistent with the PA of the torus and different from that of the galaxy disk in the central kiloparsec. Similarly, the kinematic PA in the central region of NGC 1808 is \(\approx270\arcdeg\), which is different from that of the galaxy disk.

The measured flux densities of the continuum and lines in regions that enclose the CNDs are given in Table \ref{tab:flux}. The [\ion{C}{1}]\((2\mathrm{-}1)\)/CO\((7\mathrm{-}6)\) integrated-flux ratios are generally similar or lower compared to those in local and distant galaxies observed at low resolution \citep{IRW15,Val20,Gur22,Gur23,Viz26,Xu26}.

\subsection{Line Profiles at the Galaxy Nucleus}\label{subsec:spec}

\begin{table*}[ht!]
\begin{center}
\caption{Measured Flux Densities in the CND\label{tab:flux}}
\begin{tabular}{lcc}
\tableline\tableline
 & NGC 613 (\(r<0\farcs7\)) & NGC 1808 (\(r<1\farcs5\)) \\
\tableline
\(S_{800}~[\mathrm{mJy}\)] & \(75.49\pm0.78\) & \(1079.0\pm2.5\) \\
\(\int S_\mathrm{[CI](2-1)}dv~[\mathrm{Jy~km~s}^{-1}\)] & \(165.2\pm1.4\) & \(643.8\pm3.5\) \\
\(\int S_\mathrm{CO(7-6)}dv~[\mathrm{Jy~km~s}^{-1}\)] & \(267.6\pm1.5\) & \(1176.1\pm3.5\) \\
\(\int S_\mathrm{[CI](2-1)}dv/\int S_\mathrm{CO(7-6)}dv\) & \(0.6173\pm0.0063\) & \(0.5474\pm0.0034\) \\
\tableline
\tableline
\end{tabular}
\end{center}
\tablecomments{The integrated flux densities were calculated from continuum-subtracted cubes. The uncertainties do not include the absolute calibration accuracy (\(20\%\) in Band 10 according to ALMA Cycle 10 Proposer's Guide).}
\end{table*}

\begin{figure}[ht!]
\epsscale{1}
\plotone{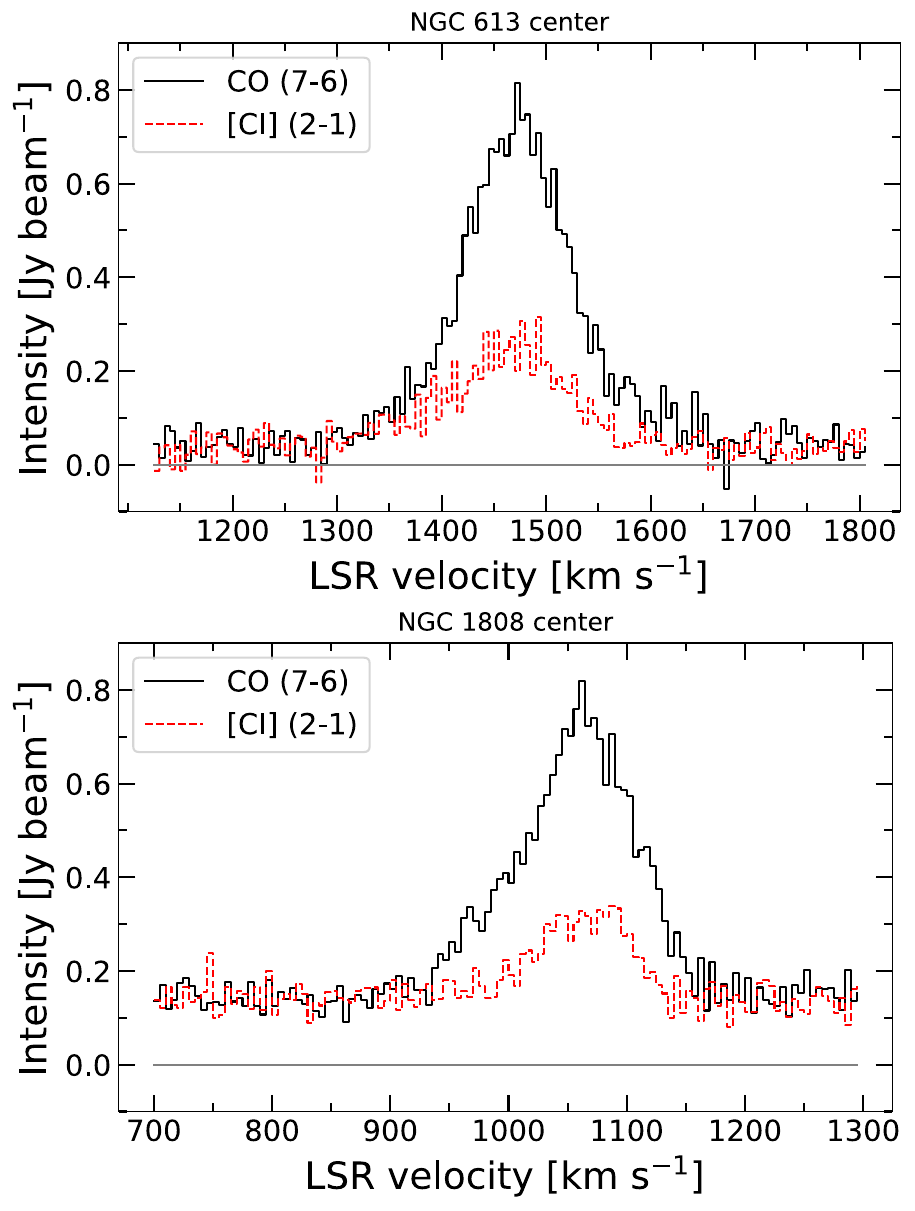}
\caption{Spectra extracted from the coordinates of the galaxy center at 10-pc resolution.\label{fig:spec}}
\end{figure}

In Figure \ref{fig:spec}, we show the [\ion{C}{1}] \((2\mathrm{-}1)\) and CO \((7\mathrm{-}6)\) spectra extracted from the position of the galaxy nucleus. The integrated intensities of the lines, calculated after subtracting the continuum, are \(\mathcal{I}_\mathrm{[CI](2-1)}=6.0\pm1.1~\mathrm{Jy~beam^{-1}~km~s^{-1}}\) and \(\mathcal{I}_\mathrm{CO(7-6)}=17.5\pm1.1~\mathrm{Jy~beam^{-1}~km~s^{-1}}\) integrated within \(v_\mathrm{LSR}=(1270,1670)~\mathrm{km~s^{-1}}\) in NGC 613, and \(\mathcal{I}_\mathrm{[CI](2-1)}=3.83\pm0.88~\mathrm{Jy~beam^{-1}~km~s^{-1}}\) and \(\mathcal{I}_\mathrm{CO(7-6)}=13.73\pm0.77~\mathrm{Jy~beam^{-1}~km~s^{-1}}\) integrated within \(v_\mathrm{LSR}=(895,1195)~\mathrm{km~s^{-1}}\) in NGC 1808. As a result, the integrated-intensity ratios are \(\mathcal{I}_\mathrm{[CI](2-1)}/\mathcal{I}_\mathrm{CO(7-6)}=0.343\pm0.066\) in NGC 613 and \(\mathcal{I}_\mathrm{[CI](2-1)}/\mathcal{I}_\mathrm{CO(7-6)}=0.279\pm0.066\) in NGC 1808. Although there is difference in angular resolution (9 and 15 pc), the two ratios are consistent with each other within the uncertainties. Note that the ratios are smaller than the integrated-flux ratios in the whole CND (Table \ref{tab:flux}) by a factor of two. This is because CO \((7\mathrm{-}6)\) is more compact than [\ion{C}{1}] \((2\mathrm{-}1)\) and concentrated in the torus.

The total line widths of [\ion{C}{1}] \((2\mathrm{-}1)\) and CO \((7\mathrm{-}6)\) are similar, but the shapes of the profiles exhibit some differences. This is especially prominent in NGC 1808, where the spectrum of [\ion{C}{1}] is ``flattened'' on the top, whereas CO exhibits a Gaussian-like peak. This indicates a complex substructure with multiple components that can be caused by variations in physical conditions, optical depth, and C/CO abundance ratio. The differences are largely smoothed out when the resolution is lowered, implying that large-scale motion of the gas in the CND is not much different between tracers. This allows us to model the gas excitation assuming a single component in Section \ref{sec:nlte} (see Figure \ref{fig:radex2}). However, we will return to the line profiles in Section \ref{sec:dyn} where we discuss gas dynamics in the torus and its surroundings.

Figure \ref{fig:spec} also shows that the intensity of the 800-GHz continuum is a significant fraction of the line intensity in NGC 1808 at this angular resolution. If the continuum is optically thin and the line is optically thick, the continuum-subtracted integrated intensity may be an underestimate of its true intensity. A similar effect can arise if the dust continuum is optically thick (\(\tau_\mathrm{d}\gtrsim1\)) and the observed line-to-continuum contrast becomes reduced (e.g., \citealt{Pap10}). This might affect the observed lines at high frequencies toward dust-enshrouded regions.

\section{Gas Physical Conditions}\label{sec:gpc}

In this section, we derive the physical conditions (density, temperature and C/CO abundance ratio) using all [\ion{C}{1}] and CO line data that are available at the resolution that resolves the CND. In Section \ref{sec:lte}, we first assume local thermodynamic equilibrium (LTE) and optically thin emission and use only [\ion{C}{1}] lines to obtain a lower limit for the column density of C. The [\ion{C}{1}] \((1\mathrm{-}0)\) line is expected to be moderately optically thin in warm gas as supported by observations of Galactic star-forming clouds \citep{ZBG86,WS95,Ike99,Ike02} and extragalactic systems (e.g., \citealt{Wei03}), albeit can be opaque in dark clouds \citep{Tat99,Mae99}. On the other hand, the [\ion{C}{1}] \((1\mathrm{-}0)\) line becomes nearly thermalized if the gas density is sufficiently high (\(\gtrsim10^4~\mathrm{cm}^{-3}\)), i.e., higher than the critical density. Similarly, we assume in this section that [\ion{C}{1}] \((2\mathrm{-}1)\) is also optically thin and thermalized.

In Section \ref{sec:nlte}, we expand the analysis to non-LTE modeling that takes into account the optical depth effects. We first analyze only [\ion{C}{1}] line ratios and then include CO lines. Both LTE and non-LTE analyses are conducted to evaluate the differences and to compare the results with previous studies. A non-LTE approach also allows us to constrain the C/CO abundance ratio simultaneously with gas density and temperature.

\subsection{[\ion{C}{1}] Line Ratio and LTE Diagnostics}\label{sec:lte}

If an emission line is optically thin, the total column density in LTE can be expressed (e.g., \citealt{MS15} and Appendix in \citealt{Sal19}) as

\begin{equation}\label{eq:cdci1}
N=\frac{8\pi k\nu_{ul}^2Z(T_\mathrm{ex})}{hc^3A_{ul}g_u}e^{E_u/{kT_\mathrm{ex}}}W_{ul}
\end{equation}
where \(W_{ul}=\int T_{\mathrm{R}}dv\) is the integrated intensity, \(T_\mathrm{R}\) is the radiation temperature, \(k\) is the Boltzmann constant, \(h\) is the Planck constant, \(c\) is the speed of light in vacuum, \(A_{ul}\) is the Einstein coefficient for spontaneous emission in a transition \(u\rightarrow l\), \(\nu_{ul}\) is the frequency of the transition, \(g_u\) is the statistical weight of the upper state, \(E_u\) is the energy of the state, \(T_\mathrm{ex}\) is the excitation temperature, and \(Z\) is the partition function. In the case of the fine structure of the carbon atom, \(g_1=3\), \(g_2=5\), \(E_1/k=23.620~\mathrm{K}\), \(E_2/k=62.462~\mathrm{K}\), and \(Z=1+g_1e^{-E_1/kT_\mathrm{ex}}+g_2e^{-E_2/kT_\mathrm{ex}}\). \(T_\mathrm{R}~[\mathrm{K}]\) is proportional to the intensity \(I_\nu\) \([\mathrm{Jy~beam}^{-1}]\) via the Rayleigh-Jeans formula, \(T_\mathrm{R}=1.222\times10^6I_\nu(\nu^2\theta_\mathrm{maj}\theta_\mathrm{min})^{-1}\), where \(\nu\) is in GHz, and \(\theta\) is the beam size (maj = major axis, min = minor axis) in arcseconds.

As \(W_{ul}\) is measured at the telescope, the only unknown in Equation (\ref{eq:cdci1}) is \(T_\mathrm{ex}\). In LTE, the excitation temperatures of all energy levels are equal, hence the ratio of integrated intensities is

\begin{equation}
R_\mathrm{[CI]}\equiv\frac{W_{21}}{W_{10}}=\frac{g_2A_{21}\nu_{10}^2}{g_1A_{10}\nu_{21}^2}e^{-\Delta E/{kT_\mathrm{ex}}},
\end{equation}
where \(\Delta E=E_2-E_1\). For \(A_{10}=7.88\times10^{-8}~\mathrm{s}^{-1}\) and \(A_{21}=2.65\times10^{-7}~\mathrm{s}^{-1}\), the excitation temperature is

\begin{equation}\label{eq:tex}
T_\mathrm{ex}=\frac{\Delta E}{k}\left[\ln\left(\frac{g_2A_{21}\nu_{10}^2}{g_1A_{10}\nu_{21}^2}\frac{1}{R_\mathrm{[CI]}}\right)\right]^{-1}=\frac{38.8~\mathrm{K}}{\ln(2.07/R_\mathrm{[CI]})}.
\end{equation}

Once \(T_\mathrm{ex}\) is calculated, we can evaluate Equation (\ref{eq:cdci1}) to get the column density. In the case of [\ion{C}{1}] \((1\mathrm{-}0)\), the column density is

\begin{equation}\label{eq:cdci2}
N_\mathrm{C}=1.99\times10^{15}Z(T_\mathrm{ex})e^{23.62~\mathrm{K}/T_\mathrm{ex}}W_{10},
\end{equation}
where \(N_\mathrm{C}\) is in \(\mathrm{cm^{-2}}\), and \(W_{10}\) is in \(\mathrm{K~km~s^{-1}}\). The value of \(N_\mathrm{C}\) given by Equation (\ref{eq:cdci2}) only weakly depends on \(T_\mathrm{ex}\) at \(T_\mathrm{ex}\gtrsim20~\mathrm{K}\) (e.g., \citealt{Wei05,Sal19}). This makes [\ion{C}{1}] lines robust probes of column density in relatively warm gas as long as LTE holds and lines are optically thin. If the lines are optically thick, the column density derived using Equation (\ref{eq:cdci2}) is a lower limit. In addition, the excitation temperature in LTE is equal to the gas kinetic temperature (\(T_\mathrm{ex}=T_\mathrm{k}\)), hence Equation (\ref{eq:tex}) can serve as a probe of gas temperature. However, at densities comparable to or lower than the critical density, the excitation is sub-thermal (\(T_\mathrm{ex}<T_\mathrm{k}\)) and non-LTE modeling is necessary.

If the optical depth of the dust continuum is non-negligible and increases with frequency, the observed \(R_\mathrm{[CI]}\) can be suppressed compared to the intrinsic ratio \citep{PDM22}. The gas surface density that results in \(\tau_\mathrm{d}=1\) at \(800~\mathrm{GHz}\) can be estimated as follows. The optical depth is \(\tau_\mathrm{d}=\kappa_\nu\Sigma_\mathrm{d}\), where \(\kappa_\nu\) is the absorption coefficient at frequency \(\nu\), and \(\Sigma_\mathrm{d}\) is the dust surface density. The absorption coefficient can be modeled as \(\kappa_\nu=\kappa_0(\nu/\nu_0)^\beta\), where we adopt \(\kappa_0=0.15~\mathrm{m^2~kg^{-1}}\) at \(\nu_0=350~\mathrm{GHz}\), and \(\beta=1.8\) (e.g., \citealt{CNC14}). This results in \(\kappa_\nu=0.66~\mathrm{m^2~kg^{-1}}\) at \(\nu=800~\mathrm{GHz}\). The dust surface density that yields \(\tau_\mathrm{d}=1\) is then \(\Sigma_\mathrm{d}=1/\kappa_\nu=1.5~\mathrm{kg~m}^{-2}\), i.e., \(\Sigma_\mathrm{gas}\sim7.2\times10^4~M_\sun~\mathrm{pc}^{-2}\) for a gas-to-dust ratio of 100. The corresponding column density of hydrogen is \(N_\mathrm{H}\sim6\times10^{24}~\mathrm{cm}^{-2}\). Compared to the column densities of H\(_2\) obtained from previous CO observations, it is likely that \(\tau_\mathrm{d}<1\) over most of the CNDs of our targets, though large opacities of \(\tau_\mathrm{d}\sim1\) toward the tori cannot be ruled out. Given the uncertainty in spatial and frequency variation of \(\tau_\mathrm{d}\), we do not include the dust opacity effects in the analysis, but note that if corrected for the above factor, \(T_\mathrm{ex}\) may be higher. However, because of the very weak dependence on \(T_\mathrm{ex}\) at these temperatures, \(N_\mathrm{C}\) increases by only \(\sim10\%\) when \(T_\mathrm{ex}\) is increased by a factor of two.

\begin{figure*}[ht!]
\epsscale{1}
\plotone{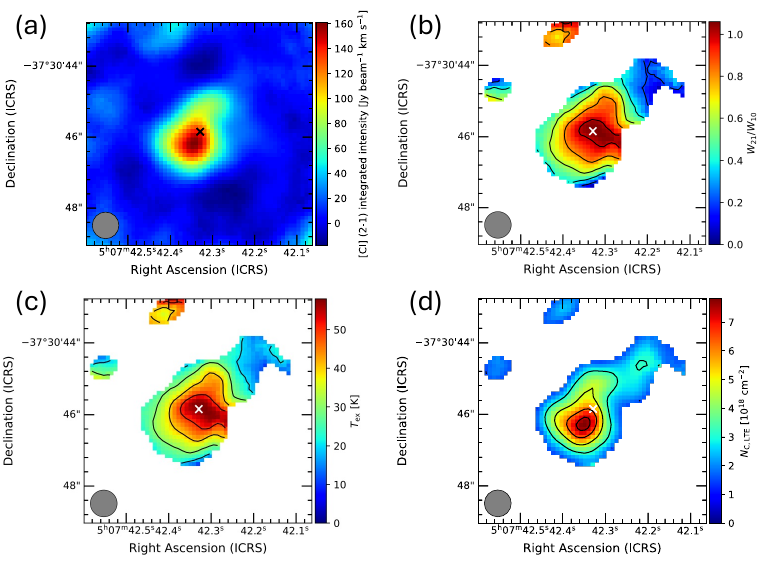}
\caption{\emph{(a)} Moment zero image of [\ion{C}{1}] \((2\mathrm{-}1)\) in the CND of NGC 1808 smoothed to a circular beam \(0\farcs75\) and produced with a threshold of \(-3\sigma\). \emph{(b)} Ratio \(R_\mathrm{[CI]}\equiv W_{21}/W_{10}\), where \(W\) is the integrated intensity in \(\mathrm{K~km~s^{-1}}\). The contours are plotted at 0.2, 0.4, 0.6, 0.8, 1.0. \emph{(c)} Excitation temperature with contours at 10, 20, 30, 40, 50 K; the peak value is 57.8 K. \emph{(d)} Column density calculated using Equation (\ref{eq:tex}) with contours at 0.2, 0.4, 0.6, 0.8, 0.95 of the maximum \(2.87\times10^{18}~\mathrm{cm}^{-2}\). The maps are masked below 20\% (\(\approx5\sigma\)) of the peak integrated intensity of [\ion{C}{1}] \((1\mathrm{-}0)\). The galaxy nucleus is indicated by the \textsf{x}-symbol.\label{fig:lte}}
\end{figure*}

\subsubsection{NGC 1808}

We performed the analysis described above using the available [\ion{C}{1}] \((1\mathrm{-}0)\) and \((2\mathrm{-}1)\) data to get the distributions of \(T_\mathrm{ex}\) and \(N_\mathrm{C}\) in the CND of NGC 1808. The results are shown in Figure \ref{fig:lte}. The data were smoothed to a common circular beam of \(0\farcs75\) (\(\approx39~\mathrm{pc}\)), which is close to the original beam size of the [\ion{C}{1}] \((1\mathrm{-}0)\) cube. Figure \ref{fig:lte}(a) shows that the peak integrated intensity of [\ion{C}{1}] \((2\mathrm{-}1)\) of the smoothed image is offset from the position of the galaxy center. This is caused by the spiral structure that extends southeast of the center (Figure \ref{fig:ci21}). However, the integrated intensity ratio \(R_\mathrm{[CI]}\) at this angular resolution clearly exhibits a peak of \(1.06\) at the galaxy nucleus. The mean value in the CND is \(0.61\) with a standard deviation of \(0.26\). The derived excitation temperature also peaks at the center, where the value is \(57.9~\mathrm{K}\), which is a factor of two higher than the mean value. The peak column density is found to be \(N_\mathrm{C}\approx7.8\times10^{18}~\mathrm{cm}^{-2}\) with an uncertainty of about 20\%. Since the derivation is correct for optically thin emission, the estimated \(N_\mathrm{C}\) is a lower limit, and may be higher by a factor of \(\tau/(1-e^{-\tau})\). For an abundance ratio of \(\mathrm{C/H}\sim10^{-4}\), this implies a hydrogen column density of \(N_\mathrm{H}\sim10^{23}~\mathrm{cm}^{-2}\).

\subsubsection{NGC 613}

We did not perform a two-dimensional analysis for NGC 613 because the [\ion{C}{1}] \((1\mathrm{-}0)\) image exhibits a non-negligible (\(0\farcs07\)) positional offset compared to other data. This discrepancy was noticed by \citet{Miy18}, who suggested that it was caused by an error in phase calibration. We found that the peaks of [\ion{C}{1}] \((1\mathrm{-}0)\) and \((2\mathrm{-}1)\) integrated-intensity images also have a similar offset as reported previously. Thus, we took a ratio of the mean values from within one beam centered at the brightest pixel of each integrated-intensity image assuming that the peaks are spatially coincident. To compare with NGC 1808 at the same linear scale, the resolution was set to \(0\farcs463\) (\(\approx39~\mathrm{pc}\)). The resulting beam-averaged integrated-intensity ratio and excitation temperature in NGC 613 were found to be \(R_\mathrm{[CI]}=1.10\) and \(61.3~\mathrm{K}\), respectively. The [\ion{C}{1}] \((1\mathrm{-}0)\) integrated intensity then yields a column density of \(N_\mathrm{C}\approx6.5\times10^{18}~\mathrm{cm}^{-2}\). By comparison, applying the same method to the center of NGC 1808, the beam-averaged values are \(R_\mathrm{[CI]}=0.93\) and \(48.5~\mathrm{K}\). Thus, the excitation temperature under LTE and optically thin assumption is a factor of \(\approx1.3\) higher at the nucleus of NGC 613.

\subsubsection{Comparison with Previous Observations}

Although the imaged spatial scale is much different, we compare the measured \(R_\mathrm{[CI]}\) with previous low-resolution observations that sampled both [\ion{C}{1}] lines in extragalactic sources. The ratio of \(R_\mathrm{[CI]}=1.0\) is high compared to that found in the central regions of nearby LIRGs (\(\approx0.4\)) \citep{Jia17}, ULIRGs (\(0.80\pm0.05\)) and starbursts (\(0.69\pm0.04\)) \citep{IRW15}, and even higher compared to the disks of nearby star-forming galaxies at 1 kpc scale (median value \(0.29\pm0.09\)) \citep{Jia19}. The excitation temperature is also high compared to nearby galaxy disks where \(T_\mathrm{ex}=20.2\pm4.2~\mathrm{K}\) with a maximum at \(\approx35~\mathrm{K}\) \citep{Jia19,Cro19,Miy21}. Similarly, \(R_\mathrm{[CI]}\) is higher than the galaxy-scale value of \(\approx0.5\) in a sample of submillimeter galaxies and quasars at redshift \(z\approx2\mathrm{-}4\) \citep{Wei03,Dan11,Wal11,AZ13,Bri19,Nes19}. The mean excitation temperature in these high-redshift galaxies is \(T_\mathrm{ex}=29.1\pm6.3~\mathrm{K}\), little higher than in local disk galaxies. \citet{Gur23} obtained \(34.5\pm2.1~\mathrm{K}\) for a sample of lensed dusty star-forming galaxies at similar redshifts. \citet{Val20} compiled a sample of both local and distant galaxies (excluding AGN) and reported a mean value of \(T_\mathrm{ex}=25.6\pm1.0~\mathrm{K}\). Compared to these previous observations at much lower resolution, the line ratios and excitation temperatures in the CNDs of NGC 613 and NGC 1808 are a factor of two higher. Our high-resolution observations clearly show that \(R_\mathrm{[CI]}\) and \(T_\mathrm{ex}\) are enhanced in the nuclei of Seyfert galaxies.

\subsection{Gas Conditions and C/CO Abundance Ratio}\label{sec:nlte}

Previous works have pointed out that the LTE approximation may not be valid for [\ion{C}{1}] in various ISM environments (e.g., \citealt{PDM22}). For instance, the [\ion{C}{1}] lines deviate from LTE at low gas densities (\(n_\mathrm{H_2}\lesssim10^3~\mathrm{cm}^{-3}\)), resulting in sub-thermal excitation (e.g., \citealt{Glo15}). On the other hand, the optical depth can be \(\tau\sim1\), especially toward regions with high column densities (\(N_\mathrm{C}\sim10^{18}~\mathrm{cm}^{-2}\)) in cold clouds. Moreover, there is no simple way to estimate the volume density of the gas and the C/CO abundance ratio, because CO lines are optically thick.

If the lines are thermalized and optically thick, the brightness temperature is equal to the kinetic temperature at the surface of a molecular cloud and can be used as a probe of the gas temperature if the source is resolved. The blackbody brightness temperature can be expressed as \(T_\mathrm{B}=(h\nu/k)[\ln(h\nu/kT_\mathrm{R}+1)]^{-1}\), where \(T_\mathrm{R}\) is the radiation temperature measured at the telescope. For instance, the peak CO \((7\mathrm{-}6)\) intensity in NGC 613 (see Figure \ref{fig:spec}) calculated using this equation is \(T_\mathrm{R}=28~\mathrm{K}\) (\(T_\mathrm{B}=44~\mathrm{K}\)) whereas the peak [\ion{C}{1}] \((2\mathrm{-}1)\) intensity is \(T_\mathrm{R}=9~\mathrm{K}\) (\(T_\mathrm{B}=23~\mathrm{K}\)). Clearly, \(T_\mathrm{R}<T_\mathrm{B}\) because the Rayleigh-Jeans approximation is not valid. However, note that this \(T_\mathrm{B}\) is lower than \(T_\mathrm{ex}\) of [\ion{C}{1}] at the galaxy nucleus derived in Section \ref{sec:lte}. This may be because the line is optically thin, not thermalized, and/or the beam-filling factor is less than unity. It will be shown below that CO \((7\mathrm{-}6)\) is optically thick, [\ion{C}{1}] \((2\mathrm{-}1)\) is optically thin, and both lines are sub-thermally excited.

\begin{figure*}[ht!]
\epsscale{1}
\plotone{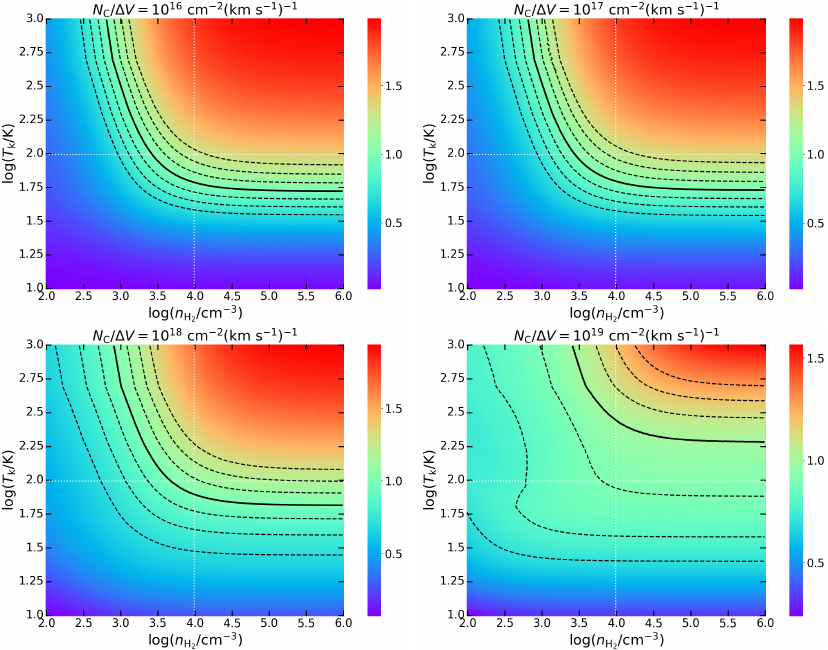}
\caption{RADEX results for the [\ion{C}{1}]\((2\mathrm{-}1)/(1\mathrm{-}0)\) line intensity ratio \(R_\mathrm{[CI]}\equiv W_{21}/W_{10}\) (color), where \(W=\int T_\mathrm{R}dv\). The contours \(R_\mathrm{[CI]}\) are plotted from 0.7 to 1.3 in steps of 0.1; the solid curve is 1.0.\label{fig:radex}}
\end{figure*}

\subsubsection{Non-LTE Modeling of [\ion{C}{1}] Line Intensities}

To further constrain the gas conditions, we used the non-LTE radiative transfer code RADEX \citep{vdT07}. The code iteratively solves statistical equilibrium equations with collisional and radiative processes and produces atomic and molecular line intensities arising from a homogeneous medium (``one zone''). The effects of optical depth are treated with a photon escape probability (\(\beta\)). The input parameters are the gas density (\(n_\mathrm{H_2}\)), kinetic temperature  (\(T_\mathrm{k}\)), column density (\(N\)), and one-dimensional line width (\(\Delta v\)). Since the optical depth in calculations is proportional to \(N/\Delta v\), the results below are presented in terms of this ratio. We ran the code for the values from \(N_\mathrm{C}/\Delta v=10^{16}\) to \(10^{19}~\mathrm{cm^{-2}(km~s^{-1})^{-1}}\) in steps of 1 dex. The lowest value corresponds to the lower limit of \(10^{18}~\mathrm{cm}^{-2}\) (optically thin approximation) and \(\Delta v=100~\mathrm{km~s}^{-1}\), which is comparable to the observed line width in the beam. \(N_\mathrm{C}/\Delta v\) is not likely to exceed \(10^{19}\) because it would require a very high C/H abundance (\(>10^{-3}\)) or a very small line width (\(\sim1~\mathrm{km~s}^{-1}\)). We assume that the line widths and beam-filling factors are equal for all lines, that the collision partners for C atoms are mainly H\(_2\) molecules, and adopt the escape probability of a large velocity gradient given as \(\beta=(1-e^{-\tau})/\tau\) (see \citet{vdT07} and the references within).

The results of calculations for the two [\ion{C}{1}] lines are shown in Figure \ref{fig:radex}. It is clear that the observed integrated-intensity ratios of \(R_\mathrm{[CI]}\approx1.0\) constrain the gas density and kinetic temperature to above \(n_\mathrm{H_2}\gtrsim10^3~\mathrm{cm}^{-3}\) and \(T_\mathrm{k}>50~\mathrm{K}\) in all models. However, there is a degeneracy between the two parameters, and the column density of C is not well constrained. To probe the C/CO abundance ratio, we add CO lines to the modeling in the following section.

\begin{table}[ht!]
\begin{center}
\caption{Integrated Intensities in the CND}\label{tab:rat}
\begin{tabular}{lcc}
\tableline\tableline
Galaxy & NGC 613 & NGC 1808 \\
\tableline
\(W~[\mathrm{K~km~s^{-1}}]\) \\
\(\mathrm{[CI]}\) \((1\mathrm{-}0)\) & \(249.2\pm7.2\) & \(323.8\pm3.1\) \\
\(\mathrm{[CI]}\) \((2\mathrm{-}1)\) & \(259.7\pm5.4\) & \(290.9\pm2.6\) \\
CO \((1\mathrm{-}0)\) & \(913.9\pm4.3\) & - \\
CO \((2\mathrm{-}1)\) & \(1302.4\pm4.3\) & \(1609.2\pm1.3\) \\
CO \((3\mathrm{-}2)\) & \(1381.9\pm1.8\)  & \(1439.6\pm2.2\)  \\
CO \((7\mathrm{-}6)\) & \(528.5\pm4.3\)  & \(630.1\pm2.3\) \\
\tableline
\(W\) Ratios \\
\(\mathrm{[CI]}\)\((2\mathrm{-}1)/(1\mathrm{-}0)\) & \(1.04\pm0.23\)  & \(0.90\pm0.14\) \\
\(\mathrm{[CI]}\)\((1\mathrm{-}0)\)/CO\((2\mathrm{-}1)\) & \(0.191\pm0.038\) & \(0.201\pm0.030\) \\
CO\((3\mathrm{-}2)/(2\mathrm{-}1)\) & \(1.06\pm0.15\) & \(0.89\pm0.13\) \\
CO\((7\mathrm{-}6)/(2\mathrm{-}1)\) & \(0.406\pm0.060\) & \(0.392\pm0.056\)  \\
\tableline
\end{tabular}
\end{center}
\tablecomments{Line intensities integrated over the velocity range shown in Figure \ref{fig:radex2} (top) with resolution 5 km s\(^{-1}\) within one beam at the center. The uncertainties of the \(W\) ratios include noise and calibration errors (10\%) added in quadrature. The spectrum of [\ion{C}{1}] (1--0) in NGC 613 is extracted from a beam centered at \((\alpha,\delta)_\mathrm{ICRS}=(01^\mathrm{h}34^\mathrm{m}18\fs1938,-29\arcdeg25\arcmin06\farcs603)\).}
\end{table}

\subsubsection{Simultaneous Non-LTE Modeling of [\ion{C}{1}] and CO}\label{sec:nltecico}

\begin{figure*}[ht!]
\epsscale{1}
\plotone{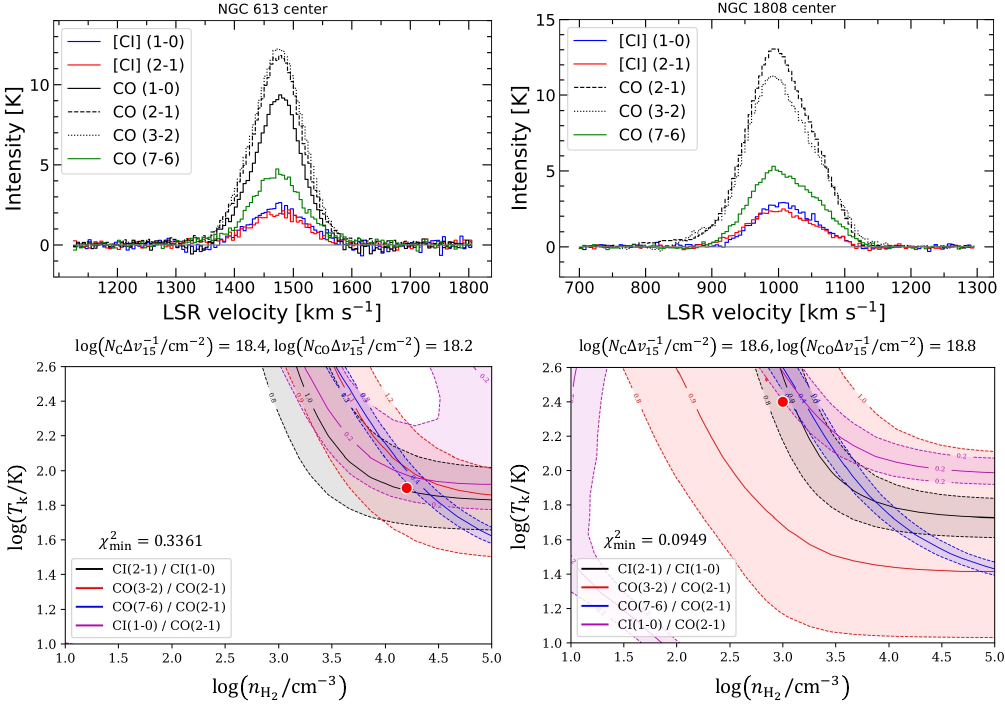}
\caption{\emph{Top.} Spectra extracted from a circular beam centered at the coordinates of the galaxy center in NGC 613 (left; resolution \(0\farcs74\)) and NGC 1808 (right; resolution \(1\farcs2\)) at the common physical resolution of \(\approx63~\mathrm{pc}\). The intensity is \(T_\mathrm{R}~[\mathrm{K}]\). \emph{Bottom.} \(\chi^2_\mathrm{min}\) solutions for [\ion{C}{1}] and CO lines in NGC 613 (left) and NGC 1808 (right). \(\chi^2_\mathrm{min}\) is indicated by a red circle. The column densities are shown as \(\log(N\Delta v_{15}^{-1}/\mathrm{cm}^{-2})\) where \(\Delta v_{15}\equiv\Delta v/(15~\mathrm{km~s^{-1}})\).\label{fig:radex2}}
\end{figure*}

To model [\ion{C}{1}] and CO line intensities, we collected all data at comparable resolution and sensitivity and reduced them following the method described in Section \ref{subsec:arc}.  All images were convolved to a common circular beam: \(0\farcs74\) for NGC 613 and \(1\farcs2\) for NGC 1808, where the resolution was limited by the archival data. The physical scale in both galaxies was matched to \(\approx63~\mathrm{pc}\). The beam-averaged line intensities measured within one beam centered at the galaxy center are given in Table \ref{tab:rat} and the spectra are shown in Figure \ref{fig:radex2}. The intensities are expressed as \(T_\mathrm{R}~\mathrm{[K]}\) to directly compare with RADEX solutions.

The parameters to be constrained are \(n_\mathrm{H_2}\), \(T_\mathrm{k}\), and \(N_\mathrm{C}/N_\mathrm{CO}\). Note that if the line widths and beam-filling factors of [\ion{C}{1}] and CO are equal, \(N_\mathrm{C}/N_\mathrm{CO}\) and hence the C/CO abundance ratio can be obtained even if the column densities of C and CO remain uncertain. It is assumed that the medium is homogeneous and that the main collision partner with C atoms and CO molecules is H\(_2\). Although the structure of the torus and its surroundings is likely much more complex (Sections \ref{sec:res} and \ref{sec:dyn}), and C/CO is not uniform (e.g., \citealt{Wad12,WSM16}), simulations suggest that most of the [\ion{C}{1}] and CO emission arises from a common disk \citep{Izu18}. The lines used in the calculations are listed in Table \ref{tab:rat}. CO \((1\mathrm{-}0)\) data exist for both galaxies, but we did not use them because the angular resolution for NGC 1808 was too low (\(\approx2''\)) and we wanted to use the same pairs of lines for both galaxies. Only \(^{12}\)CO data were used, and not those of \(^{13}\)CO and C\(^{18}\)O to avoid having the abundances and beam-filling factors of isotopologues as additional free parameters.

We first ran the code to get the line intensities in each grid point of the parameter space, and then calculated

\begin{equation}\label{eq:chi}
\chi^2(\alpha)=\sum_i\frac{[R_i^\mathrm{mod}(\alpha)-R_i^\mathrm{obs}(\alpha)]^2}{\sigma_i^2},
\end{equation}
where \(\alpha=(n_\mathrm{H_2},T_\mathrm{k},N_\mathrm{C}/\Delta v,N_\mathrm{CO}/\Delta v)\) is a set of parameters, \(R\) is the ratio of integrated intensities, and \(\sigma\) is the uncertainty. The summation has a term for each line integrated-intensity ratio that was used in the analysis (four in total). We fixed the line width to \(\Delta v=15~\mathrm{km~s}^{-1}\) and varied \(N\). This line width is comparable to those observed toward molecular clumps in the Galactic center (e.g., \citealt{MT00}), but the exact value used here is not important because the calculations depend on the ratio \(N/\Delta v\). Note that \(\chi^2\) was calculated using the integrated-intensity ratios, rather than the integrated intensities themselves. This was done in order to avoid having the beam-filling factor as an additional free parameter. It was assumed that the beam-filling factor is the same for all lines. The uncertainty was calculated as \(\sigma=\sqrt{\sigma_\mathrm{noise}^2+\sigma_\mathrm{calib}^2}\), where \(\sigma_\mathrm{calib}\) was adopted to be 10\% for each line. Although the absolute calibration accuracy is only 20\% in Band 10, we decided to use 10\% here because all lines were measured using the same telescope with similar methods, hence relative errors are smaller.

The solutions obtained from \(\chi^2_\mathrm{min}\) for the central 63 pc of NGC 613 and NGC 1808 are shown in Figure \ref{fig:radex2} (lower panels). The gas density and kinetic temperature at \(\chi^2_\mathrm{min}\) were found to be \((n_\mathrm{H_2},T_\mathrm{k})=(10^{4.2}~\mathrm{cm}^{-3},80~\mathrm{K})\) in NGC 613 and \((n_\mathrm{H_2},T_\mathrm{k})=(10^{3.0}~\mathrm{cm}^{-3},250~\mathrm{K})\) in NGC 1808. The [\ion{C}{1}] lines are nearly thermalized (\(T_\mathrm{ex}\approx T_\mathrm{k}\)) at \(n_\mathrm{H_2}\sim10^4~\mathrm{cm}^{-3}\), but become sub-thermally excited at \(n_\mathrm{H_2}\sim10^3~\mathrm{cm}^{-3}\), and the deviation is especially prominent for [\ion{C}{1}] \((2\mathrm{-}1)\). Both lines are optically thin (\(\tau\approx0.2\mathrm{-}0.5\)). As a result, the derived column densities are comparable to those in Section \ref{sec:lte}. For a line width of \(\Delta v=50~\mathrm{km~s}^{-1}\), the obtained column density of atomic carbon in the CND of NGC 613 is \(N_\mathrm{C}\approx8.4\times10^{18}~\mathrm{cm}^{-2}\) and that of CO is \(N_\mathrm{CO}\approx5.3\times10^{18}~\mathrm{cm}^{-2}\). The C/CO abundance ratio averaged over a beam of 63 pc centered at the position of the AGN is \(\approx1.6\) in NGC 613 and \(\approx0.63\) in NGC 1808.

Note that although the observed emission within the beam of 63 pc is dominated by the bright torus, it also includes a part of the surrounding CND. Consequently, if the physical size of the torus is different in the two galaxies, the physical conditions derived here may be affected by different filling factors of the torus and the surrounding medium within the beam. Based on high-resolution CO observations, the radii of molecular tori are reported to be \(\approx8~\mathrm{pc}\) in NGC 613 and \(6\pm2~\mathrm{pc}\) in NGC 1808 \citep{Com19,Com26}.

We also performed RADEX calculations as described above using NGC 613 data at a factor of two higher angular resolution of \(0\farcs4\) (\(\approx34~\mathrm{pc}\)) to investigate the effect of angular resolution on the derived physical conditions. There are two main differences in the result obtained at higher resolution: the C/CO abundance ratio increases to \(\approx2.0\), and the density at \(\chi_\mathrm{min}^2\) increases to \(10^{4.4}~\mathrm{cm}^{-3}\). The result indicates that C/CO is enhanced in the torus compared to the rest of the CND. The ratio may be even higher at smaller spatial scale, though observations at higher angular resolution are needed to test this.

The derived values of C/CO are one order of magnitude higher than the average in Galactic clouds (\(\approx0.1\); e.g., \citealt{Ike99}), and comparable to those reported for the CND in the Galactic center (\(\approx0.3\mathrm{-}2\); e.g., \citealt{Tan11,TNK21}). However, the density and temperature of the [\ion{C}{1}]-emitting gas are higher than those reported for the Galactic CND \citep{TNK26}. The abundances are also similar to those reported for the nuclei of some nearby galaxies \citep{IB01,IB02,IB03,Hit08}, including NGC 253 \citep{Kri16}, and molecular clouds in the Small Magellanic Cloud \citep{RT16} where non-LTE radiative transfer models have yielded solutions for an enhanced C/CO. An interesting example is NGC 7469, where C/CO is estimated to be well above unity based on observations of [\ion{C}{1}] \((1\mathrm{-}0)\) and CO lines \citep{Izu20,Liu23}. This galaxy hosts a Type-1 AGN and its luminosity within \(2\mathrm{-}10~\mathrm{keV}\) is \(\log(L_\mathrm{X}/\mathrm{erg~s^{-1}})=43.2\), two orders of magnitude higher than that in NGC 613. However, most of the previous studies used fewer lines and/or at lower resolution. In this work, we have demonstrated that the abundance of C is enhanced relative to that of CO in the nuclei of Seyfert galaxies, for the first time at a resolution of 30-60 pc using both [\ion{C}{1}] lines and multiple CO lines, all taken by ALMA with matched angular and spectral resolution.

\begin{figure}[ht!]
\epsscale{1.2}
\plotone{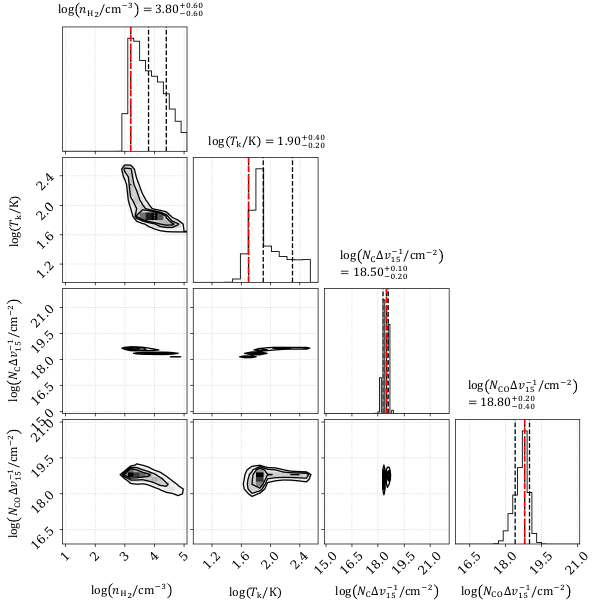}
\caption{Probability distributions for the nucleus (central pixel) of NGC 1808. The red dashed line is the most probable value, and the black dashed lines are the 16th, 50th (median; displayed values), and 84th percentile. \(\Delta v_{15}\equiv\Delta v/(15~\mathrm{km~s^{-1}})\).\label{fig:radex3}}
\end{figure}

\begin{figure*}[ht!]
\epsscale{1}
\plotone{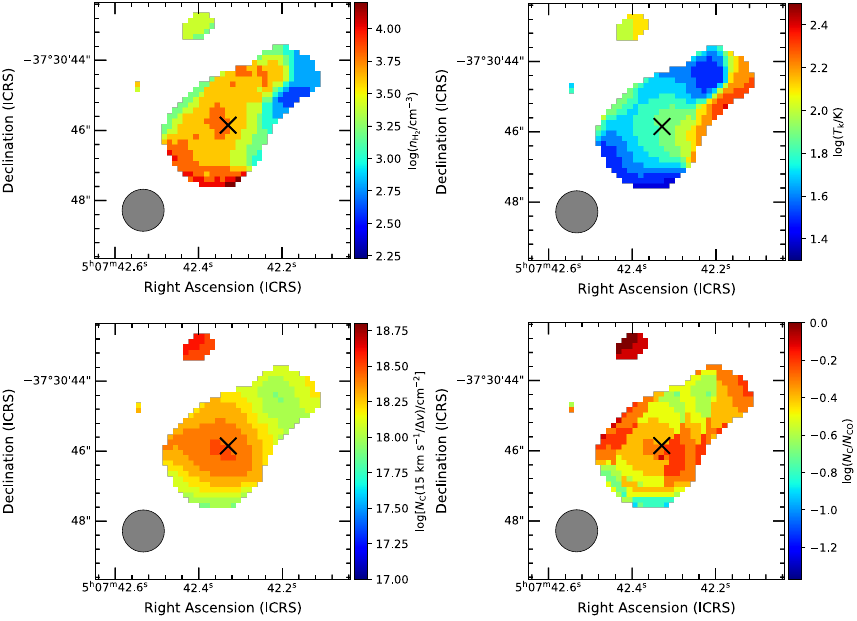}
\caption{Images of the median values of probability distributions for each parameter in NGC 1808. The pixels are masked below 20\% of the peak [\ion{C}{1}] \((1\mathrm{-}0)\) integrated intensity. The position of the galaxy nucleus is indicated by an \textsf{x}-symbol.\label{fig:radex4}}
\end{figure*}

\subsubsection{Probability Distributions}

To evaluate the accuracy of the derived physical conditions and investigate how their probabilities vary in parameter space, we calculated the probability distribution following the method presented in \citet{Ten22}. Assuming that the parameters obey a multivariate Gaussian distribution, the probability can be expressed as

\begin{equation}\label{eq:pro}
P(\alpha)\propto e^{-\chi^2(\alpha)/2},
\end{equation}
where \(\alpha\) is a set of parameters. We calculated \(P\) for each pixel in the CND of NGC 1808 using the publicly available code from \citet{Ten22} after modifying it. The result for the central pixel (galaxy nucleus) is presented in Figure \ref{fig:radex3}. The median values of \(n_\mathrm{H_2}\), \(T_\mathrm{k}\), \(N_\mathrm{C}/\Delta v\), and \(N_\mathrm{CO}/\Delta v\) are shown above each plot of one-dimensional probability distributions, whereas two-dimensional probability distributions are shown for all pairs of parameters in the plots with shaded areas. For instance, the result shows that the column density of C (divided by the line width) is the best constrained parameter at the galaxy nucleus with the median value of \(N_\mathrm{C}=10^{18.5^{+0.10}_{-0.20}}~\mathrm{cm}^{-2}(\Delta v/15~\mathrm{km~s}^{-1})\). The column density of CO is also reasonably well constrained. The distributions of column densities for C and CO imply that the C/CO abundance ratio is likely less than unity, which is consistent with the results from \(\chi^2_\mathrm{min}\) for the CND of NGC 1808. On the other hand, the gas density and temperature show wider probability distributions within \(n_\mathrm{H_2}\sim10^{3\mathrm{-}5}~\mathrm{cm}^{-3}\) and \(T_\mathrm{k}\sim40\mathrm{-}300~\mathrm{K}\) with median values of \(\approx10^{3.8}~\mathrm{cm}^{-3}\) and \(\approx79~\mathrm{K}\). Under these physical conditions, the CO \((7\mathrm{-}6)\) line is optically thick and both [\ion{C}{1}] lines are optically thin. In addition, both [\ion{C}{1}] \((2\mathrm{-}1)\) and CO \((7\mathrm{-}6)\) are sub-thermally excited. The median values for NGC 1808 here are similar to those obtained by the \(\chi_\mathrm{min}^2\) method for the nucleus of NGC 613.

In Figure \ref{fig:radex4}, we show maps of the median values of each parameter in the CND of NGC 1808 calculated using Equation (\ref{eq:pro}). The images reveal that \(n_\mathrm{H_2}\), \(N_\mathrm{C}\), and \(N_\mathrm{C}/N_\mathrm{CO}\) exhibit peaks near the position of the galaxy nucleus. The distribution of \(T_\mathrm{k}\) is similar to that of \(T_\mathrm{ex}\) derived using LTE (Figure \ref{fig:lte}), but the values are higher. High temperatures of \(T_\mathrm{k}\gtrsim100~\mathrm{K}\) are found west of the galaxy center. This region also exhibits high line ratios of [\ion{Fe}{2}] 1.64 \(\micron\)/Br\(\gamma\) and H\(_2\) 2.12 \(\micron\)/Br\(\gamma\), suggesting shock excitation \citep{Bus17}. In addition to stellar feedback, noncircular gas motions seen in H\(_2\) as well as CO and [\ion{C}{1}] velocity maps could cause gas collisions that trigger shocks and elevate the gas kinetic temperature.

\begin{figure*}[ht!]
\epsscale{1}
\plotone{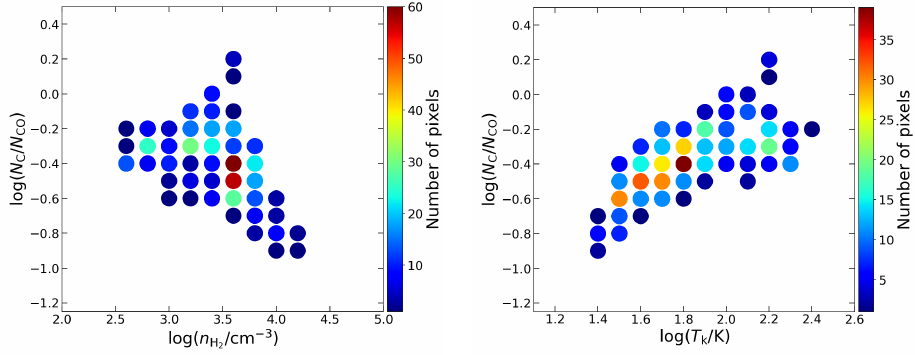}
\caption{Scatter plots of \(N_\mathrm{C}/N_\mathrm{CO}\) vs. \(n_\mathrm{H_2}\) and \(T_\mathrm{k}\) for the CND in NGC 1808 shown in Figure \ref{fig:radex4}. The color corresponds to the number of pixels.\label{fig:radex5}}
\end{figure*}

Figure \ref{fig:radex5} shows pixel-by-pixel relations between C/CO, \(T_\mathrm{k}\), and \(n_\mathrm{H_2}\). Overall, C/CO tends to increase with \(T_\mathrm{k}\) and decrease with \(n_\mathrm{H_2}\). These trends could be a result of CO dissociation in regions that are less shielded in low-density environments, where FUV and/or X-ray radiation fields that heat the gas play a key role in regulating the abundances, as expected in theoretical models of PDRs and XDRs (e.g., \citealt{WVC22}). In the next section, we discuss on the possible origin of the enhancement of C/CO in the CND.

\subsection{Origin of C/CO Enhancement in the CND}

The ISM environment in the CND is expected to be strongly affected by AGN and stellar feedback, including intense radiation field (FUV and X-rays), CRs driven by jets and supernova explosions, and mechanical heating by shocks and turbulence. In this section, we make a simple comparison of the results of observations to predictions from theoretical models of PDRs, XDRs, and CR-dominated regions (CRDRs). 

Numerous theoretical studies have modeled the line intensities and their ratios arising from such regions using numerical methods (e.g., \citealt{PTV04,MS05,MSI06,MSI07,Mei11,HTH13,BPV15,PBZ18}). In one-dimensional models, \citet{MSI06} evaluated the line intensities at different hydrogen nucleus densities (\(n_\mathrm{H}=10^3\), \(10^4\), \(10^5~\mathrm{cm}^{-3}\)), PDR radiation fluxes (\(G_0=10^2\), \(10^3\), \(10^4\), \(10^5\)), XDR radiation fluxes (\(F_\mathrm{X}=0.16\), \(1.6\), \(16\), \(160~\mathrm{erg~s^{-1}~cm^{-2}}\)), and CR ionization rates (\(\zeta_\mathrm{CR}=5\times10^{-17}\), \(5\times10^{-15}~\mathrm{s}^{-1}\)). The PDR radiation field includes FUV within \(6\mathrm{-}13.6~\mathrm{eV}\), and \(G_0=1\) corresponds to \(F_\mathrm{FUV}=1.6\times10^{-3}~\mathrm{erg~s^{-1}~cm}^{-2}\). In \citet{MSI07}, the calculations are expanded to a wider grid of parameters, whereas the effects of CRs and mechanical heating in star-forming regions are further investigated in \citet{Mei11}. The C/H abundance ratio in the gas phase is fixed to \(1.4\times10^{-4}\).

The X-ray flux at a distance of \(d=10~\mathrm{pc}\) from a point source of \(\log(L_\mathrm{X}/\mathrm{erg~s^{-1}})\sim10^{40\mathrm{-}41}\) within \(2\mathrm{-}10~\mathrm{keV}\) is \(F_\mathrm{X}=L_\mathrm{X}/(4\pi d^2)\approx0.8\mathrm{-}8~\mathrm{erg~s^{-1}~cm^{-2}}\), which is comparable to the fluxes in the models, though it must be weaker in the CND because of attenuation in the ISM. On the other hand, attenuation by dust is even stronger in FUV.

First, we compared the observed [\ion{C}{1}]\((2\mathrm{-}1)/(1\mathrm{-}0)\) ratio to the model results presented in \citet{MSI07}. The line intensity \(I~[\mathrm{erg~s^{-1}~cm^{-2}~sr^{-1}}]\), as given in their work, is proportional to the integrated intensity \(W~[\mathrm{K~km~s^{-1}}]\) as \(I=1.025\times10^{-15}\nu^3W\), where \(\nu\) is in GHz. The beam-averaged line ratio is \(I_{21}/I_{10}=(\nu_{21}/\nu_{10})^3W_\mathrm{21}/W_\mathrm{10}=4.6\pm1.0\) in NGC 613 and \(4.00\pm0.62\) in NGC 1808 at the resolution of 63 pc, with a peak of \(\approx4.9\) at the galaxy nucleus derived at higher resolution in Section \ref{sec:lte}. Assuming a gas density of \(n_\mathrm{H}=10^4~\mathrm{cm}^{-3}\) (Figure \ref{fig:radex2}), we find that a PDR with a strong radiation field (\(G_0\sim10^{3\mathrm{-}4}\)) and an XDR with a moderate/low X-ray flux (\(F_\mathrm{X}\sim0.1\mathrm{-}1~\mathrm{erg~s^{-1}~cm^{-2}}\)) from \citet{MSI07} are consistent with the observed ratios. At a lower density of \(10^3~\mathrm{cm}^{-3}\), only an XDR with \(F_\mathrm{X}\sim0.1~\mathrm{erg~s^{-1}~cm^{-2}}\) is compatible with the data.

We also compared the ratios CO \((3\mathrm{-}2)/(2\mathrm{-}1)\), CO \((7\mathrm{-}6)/(2\mathrm{-}1)\), and [\ion{C}{1}]\((1\mathrm{-}0)\)/CO\((2\mathrm{-}1)\) using the values in Table \ref{tab:rat} at the spatial scale of 63 pc to those published in \citet{MSI06}. To search for the most likely set of parameters, we calculated \(\chi^2_\mathrm{min}\) using Equation (\ref{eq:chi}) for the parameters \(n\), \(G_0\), \(F_\mathrm{X}\), and \(\zeta_\mathrm{CR}\), and each term in the sum corresponds to one pair of lines. Only [\ion{C}{1}] and \(^{12}\)CO lines were used, and not those of \(^{13}\)CO or other isotopologues, in order to mitigate the uncertainty in their abundances.

\begin{figure*}[ht!]
\epsscale{1.175}
\plotone{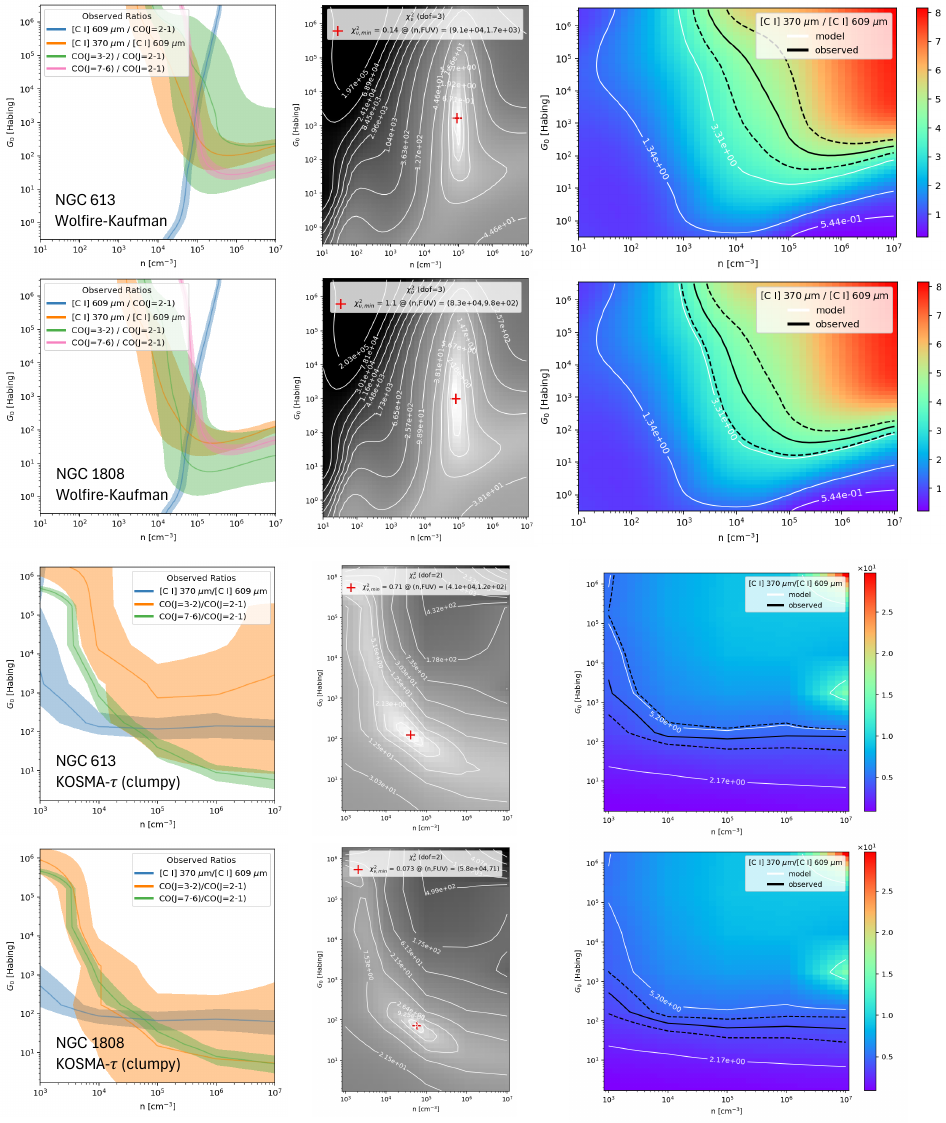}
\caption{PDRT results for the central 63-pc regions in NGC 613 and NGC 1808. Shown are the line intensity ratios (\(\mathrm{erg~s^{-1}~cm^{-2}~sr^{-1}}\) scale) (left), \(\chi^2\) maps (middle), and [\ion{C}{1}] line ratios (right) in (\(n_\mathrm{H},G_0\)) parameter space. The calculations were conducted using the Wolfire-Kaufman (upper rows) and KOSMA-\(\tau\) (lower rows) codes.\label{fig:pdr}}
\end{figure*}

The lowest values of \(\chi^2\) in the nuclei of both galaxies were obtained for \(n_\mathrm{H}=10^5~\mathrm{cm}^{-3}\), \(G_0=10^3\), and \(\zeta_\mathrm{CR}=5\times10^{-17}~\mathrm{s}^{-1}\), followed by models with higher CR rates of \(\zeta_\mathrm{CR}=5\times10^{-15}~\mathrm{s}^{-1}\). Since CR rates of \(\zeta_\mathrm{CR}\sim10^{-16}~\mathrm{s}^{-1}\) have been observed in the Galactic center region (e.g., \citealt{vdT06}), which does not harbor an AGN, values of \(\gtrsim10^{-16}~\mathrm{s}^{-1}\) are plausible. Overall, the CO line ratios in the two galaxies at the resolution of 63 pc are in better agreement with PDR models compared to XDR models. Regarding the line ratios of other diagnostic molecules, \(I_\mathrm{HCN(1-0)}/I_\mathrm{HCO^{+}(1-0)}\) is \(1.9\) in NGC 613 \citep{Miy17} and \(1.5\) in NGC 1808 \citep{Sal18}. These values are also in better agreement with PDRs than XDRs when compared to the models in \citet{MSI06}.

To further explore the possibility of a PDR and its conditions, we used the Photodissociation Region Toolbox (PDRT; \citealt{PW23}), a program that calculates the line intensity ratios produced under conditions described by physical parameters such as the hydrogen nucleus density \(n_\mathrm{H}\) and \(G_0\). We first applied the one-dimensional code Wolfire-Kaufman 2020 (``wk2020''), which is an updated version of the classical PDR \citep{TH85}, and used all [\ion{C}{1}] and CO line ratios from Table \ref{tab:rat}. The model assumes a carbon abundance of \(1.6\times10^{-4}\), CR ionization rate of \(\zeta=2\times10^{-16}~\mathrm{s}^{-1}\), and includes FUV (\(6\mathrm{-}13.6~\mathrm{eV}\)), soft X-rays (\(<0.1~\mathrm{keV}\)), and dust. The geometry is plane-parallel with constant density calculated within a range from 10 to \(10^7~\mathrm{cm}^{-3}\). We set the metallicity to the solar value and the line-of-sight angle to \(60\arcdeg\) to represent the inclination of the CND and torus.

Figure \ref{fig:pdr} (upper two rows) shows the result of calculations with ``wk2020''. The observed line ratios in both CNDs are consistent with PDRs that are remarkably similar and exhibit a high density of \(n_\mathrm{H}=(8\mathrm{-}9)\times10^4~\mathrm{cm}^{-3}\) and an FUV radiation field of \(G_0=(1\mathrm{-}2)\times10^3\). The [\ion{C}{1}] line ratios (right panels) show that the radiation field \(G_0\) increases if the density is lower than \(\sim10^5~\mathrm{cm}^{-3}\). We can also see that the minimum radiation field that can produce the observed [\ion{C}{1}] ratios at most densities is \(G_0\sim10^2\). At any fixed density, \(G_0\) is larger in NGC 613 compared to that in NGC 1808. Moreover, the minimum density for any radiation field in the calculated range is \(n_\mathrm{H}>10^3~\mathrm{cm}^{-3}\), which is consistent with the RADEX result in Figure \ref{fig:radex}.

The Wolfire-Kaufman model assumes constant density in a plane-parallel medium. However, as mentioned in Section \ref{sec:intro}, many observations of Galactic clouds indicate that C and CO are well mixed inside molecular clouds, and require clumpy PDR models to be explained. Considering the possibility that CNDs have such complex substructure, we also tried the one-dimensional code KOSMA-\(\tau\) (``kt2013wd01-25'') \citep{ROO22}. This model incorporates a clumpy medium with spherical geometry where the density is a function of clump depth. The abundance of carbon is \(2.34\times10^{-4}\) and the CR rate is \(\zeta=2\times10^{-16}~\mathrm{s}^{-1}\). We set the medium to ``clumpy'' and the maximum clump mass to \(10^3~M_\odot\).

The results of calculations are shown in Figure \ref{fig:pdr} (lower two panels). The model yields lower values of the density \(n_\mathrm{H}\) at the clump surface and radiation field \(G_0\) compared to those in ``wk2020''. As in the classical PDR model, \(G_0\) is higher in NGC 613 compared to that in NGC 1808.

Although there is some degeneracy when comparing to numerical models, the PDRT results imply that the intensities of the observed [\ion{C}{1}] and CO lines in the central 63 pc of the CNDs of both nuclei can be explained in the framework of PDRs with radiation fields \(G_0\sim10^{2-3}\). On the other hand, the [\ion{C}{1}] ratios alone are also compatible with XDR models in \citet{MSI07}. It should be noted that PDRT models do not incorporate hard X-rays, whereas the distributions of C and C/CO abundance ratio are known to be different in XDR and PDR models \citep{WVC22}.

The higher C/CO ratio in NGC 613 compared to that in NGC 1808, as derived in Section \ref{sec:nlte}, may be a result of an order of magnitude higher hard X-ray luminosity, CRs, and/or shocks associated with the jet in NGC 613. Stronger stellar feedback is unlikely because star formation efficiency is suppressed in NGC 613, as evidenced by very faint Br\(\gamma\) relative to CO \citep{FB14,Miy17}. By contrast, Pa\(\alpha\), Br\(\gamma\), and CO are all bright in the nucleus of NGC 1808, and star formation efficiency is relatively high \citep{Kot96,Bus17,Sal17}. The XDR case is supported by the observations of the CND in NGC 7469 at 100-pc resolution, whose hard X-ray luminosity is two orders of magnitude higher than in NGC 613. A very bright [\ion{C}{1}] \((1\mathrm{-}0)\) relative to low-\(J\) CO lines in this galaxy nucleus can be explained if C/CO is as high as \(\sim3\mathrm{-}10\) and the conditions are comparable to XDRs \citep{Izu20,Liu23}. On the other hand, a high C/CO of \(\sim0.7\mathrm{-}2\) is also reported for the inner part of the central molecular zone and the CND in the Galactic center, where it is likely enhanced by CRs accelerated by a nearby supernova remnant \citep{TNK21}.

Finally, there is a possibility of mechanical heating by shocks. The effects of shocks can be investigated with molecular shocked-gas tracers such as SiO \((2\mathrm{-}1)\) (e.g., \citealt{GB10}). The SiO emission is not detected at the AGN of NGC 613 \citep{Miy17}, and is only weakly detected in NGC 1808 \citep{Sal18}.  However, detections of [\ion{Fe}{2}] \(1.64~\micron\) and warm molecular hydrogen traced by the vibrational lines such as H\(_2\) \(2.12~\micron\) provide some support for the shock-heating case in the two CNDs \citep{FB14,Bus17}. While these lines can also be enhanced by radiative heating, diagnostics based on various vibrational lines of H\(_2\) suggests that the excitation of molecular hydrogen in the nuclei of the two galaxies has a significant contribution from shocks. The possible cause of shocks can be, e.g., interaction with a jet. Since the nuclear activity in the two galaxies is different, the likely origin of mechanical heating is AGN feedback in NGC 613 and stellar feedback in NGC 1808.

In conclusion, while PDR chemistry models can explain the observed [\ion{C}{1}] line intensities in NGC 613 and NGC 1808, the subtle difference in the physical conditions and C/CO abundance ratio between the two nuclei may be regulated by the ambient flux of hard X-rays, CRs, as well as mechanical heating. Further multi-line [\ion{C}{1}] and CO observations of a large sample of nuclei with a wide range of \(L_\mathrm{X}\) will provide more insight into the role of X-rays in the enrichment of atomic carbon.

\section{Gas Dynamics}\label{sec:dyn}

In this section, we analyze the line profiles to investigate the spatial origin of the emission and the dynamics of the gas traced by the observed [\ion{C}{1}] and CO lines.

\subsection{NGC 613}

\begin{figure}[ht!]
\epsscale{1}
\plotone{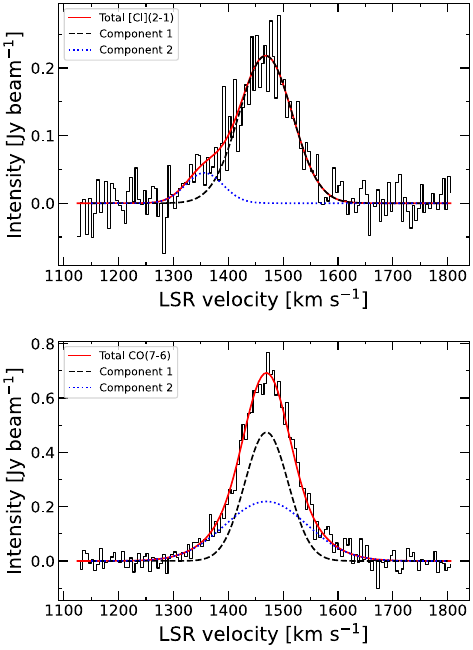}
\caption{Two-component fits of the continuum-subtracted [\ion{C}{1}] \((2\mathrm{-}1)\) and CO \((7\mathrm{-}6)\) spectra extracted from the position of the torus in NGC 613 at 15-pc resolution.\label{fig:wing}}
\end{figure}

The spectra of [\ion{C}{1}] \((2\mathrm{-}1)\) and CO \((7\mathrm{-}6)\) extracted from the position of the galaxy nucleus at high resolution (Figure \ref{fig:spec}) exhibit complex structure. To analyze the profiles, we fitted the spectra with two Gaussian functions, as shown in Figure \ref{fig:wing}. It is clear that CO \((7\mathrm{-}6)\) can be fitted with two Gaussians, where one traces the main component (torus) with a FWHM line width of \(93.3\pm9.3~\mathrm{km~s}^{-1}\) and a broad component with a line width of \(176\pm28~\mathrm{km~s}^{-1}\). As much as \(47\%\) of the CO \((7\mathrm{-}6)\) flux is found to be in the broad component. On the other hand, [\ion{C}{1}] \((2\mathrm{-}1)\) could be fitted with the main component whose FWHM line width is \(114.0\pm9.6~\mathrm{km~s}^{-1}\) and a smaller, blue-shifted one, whose fraction is only \(13\%\). The peak velocity of the blue-shifted component is \(1356\pm16~\mathrm{km~s}^{-1}\).

\begin{figure}[ht!]
\epsscale{1.15}
\plotone{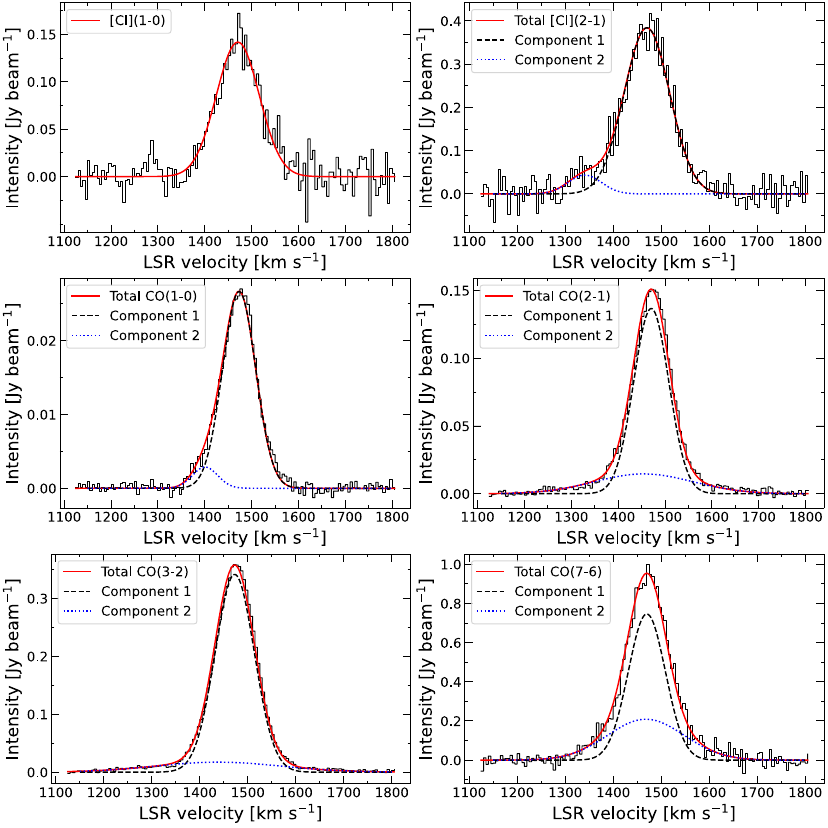}
\caption{Two-component fits of all spectra from the nucleus of NGC 613 at \(0\farcs4\) (34 pc) resolution.\label{fig:wing2}}
\end{figure}

Figure \ref{fig:wing2} shows two-component Gaussian fits on all spectra toward the nucleus of NGC 613 at a matched resolution of \(0\farcs4\) (34 pc). The secondary component is clearly detected in [\ion{C}{1}] \((2\mathrm{-}1)\) and CO \((7\mathrm{-}6)\) at this resolution as well. On the other hand, the profiles reveal even broader components in CO \((2\mathrm{-}1)\) and \((3\mathrm{-}2)\) up to \(\approx250~\mathrm{km~s}^{-1}\) with respect to the systemic velocity of \(v_\mathrm{sys}=1471~\mathrm{km~s}^{-1}\). Previous observations of CO \((3\mathrm{-}2)\) have already revealed broad lines interpreted as a molecular outflow \citep{Miy17,Aud19}. However, \citet{Com26} did not identify the outflow in their image of CO \((3\mathrm{-}2)\) at 1-pc resolution that resolves the torus, suggesting that the outflow is diffuse.

To further investigate the spatial origin of the broad component, we generated moment zero images of CO \((7\mathrm{-}6)\) and [\ion{C}{1}] \((2\mathrm{-}1)\) by integrating the cubes at velocities \(|v_\mathrm{LSR}-v_\mathrm{sys}|>120~\mathrm{km~s}^{-1}\). This velocity range is entirely dominated by the broad component.

\begin{figure}[ht!]
\epsscale{1.2}
\plotone{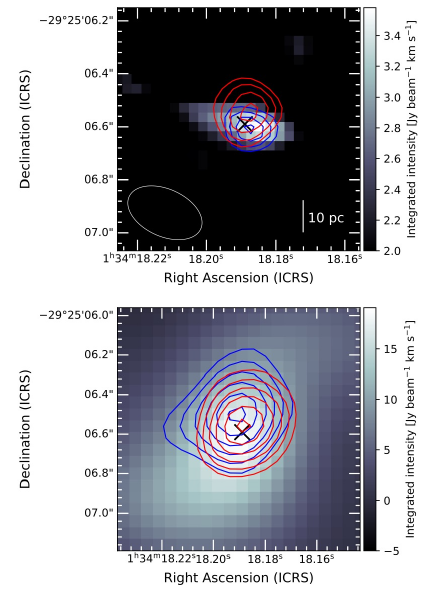}
\caption{Integrated intensity maps of the blue (\(v_\mathrm{LSR}<v_\mathrm{sys}-120~\mathrm{km~s}^{-1}\)) and red (\(v_\mathrm{LSR}>v_\mathrm{sys}+120~\mathrm{km~s}^{-1}\)) components in CO \((7\mathrm{-}6)\) (top) and CO \((2\mathrm{-}1)\) (bottom) in NGC 613 with contours at \((0.7,0.8,0.9,0.98)\times\mathcal{I}_\mathrm{CO(7-6)}^\mathrm{max}\) and \((0.5,0.6,0.7,0.8,0.9,0.98)\times\mathcal{I}_\mathrm{CO(2-1)}^\mathrm{max}\). The grey scale shows the blue component of [\ion{C}{1}] \((2\mathrm{-}1)\) (top) and CO \((2\mathrm{-}1)\) intensity integrated over all velocities (bottom). The nucleus is indicated by an \textsf{x}-symbol.\label{fig:wing3}}
\end{figure}

Figure \ref{fig:wing3} shows the integrated intensity maps of the blue-shifted and red-shifted parts of the broad component in CO \((7\mathrm{-}6)\) and CO \((2\mathrm{-}1)\) data. The maps reveal that the two lines exhibit remarkably different spatial distributions. The blue part of CO \((2\mathrm{-}1)\) at the resolution of \(0\farcs4\) is extended toward northeast, which is approximately in the direction of the large-scale jet detected at radio frequencies and an outflow seen in [\ion{O}{3}] \(\lambda5007\) \citep{Hum87,Miy17}. This spatial offset between the blue-shifted and red-shifted components is similar to that found in CO \((3\mathrm{-}2)\) images \citep{Aud19}, and is consistent with an AGN-driven outflow from the nucleus and possibly entrained by the jet. By contrast, the peak of the blue part of CO \((7\mathrm{-}6)\) at \(0\farcs2\) resolution is south of the nucleus, whereas the peak of the red part is north of it, and they are separated by \(\approx6~\mathrm{pc}\). Their orientation is consistent with the PA of the molecular torus \citep{Com26}. This result suggests that the high-velocity component in CO \((7\mathrm{-}6)\) is tracing the gas in the inner part of the torus that is rotating faster about the SMBH. Figure \ref{fig:wing3} (top) shows that the blue component of [\ion{C}{1}] \((2\mathrm{-}1)\) spatially coincides with the blue peak of CO \((7\mathrm{-}6)\). Thus, the atomic carbon gas is also distributed in the inner part of the torus. For a circular velocity of \(150~\mathrm{km~s}^{-1}\), which is the mean velocity in the blue-shifted [\ion{C}{1}] component  with respect to \(v_\mathrm{sys}\) corrected for the inclination angle of the torus (\(i=50\arcdeg\)), the mass enclosed within a radius of \(3~\mathrm{pc}\) is \(2\times10^7~M_\sun\). This is in agreement within a factor of two with the mass of the SMBH estimated by \citet{Com26} from CO \((3\mathrm{-}2)\) data.

Since [\ion{C}{1}] \((1\mathrm{-}0)\) is not clearly detected in the blue-shifted component, the line intensity ratio is \(W_\mathrm{[CI](2-1)}/W_\mathrm{[CI](1-0)}>1\) in the torus of NGC 613. In LTE approximation, such ratio yields \(T_\mathrm{ex}>50~\mathrm{K}\). Non-LTE models (Figure \ref{fig:radex}) yield that the temperature of the gas may be as high as \(T_\mathrm{k}\gtrsim100~\mathrm{K}\) for ratios significantly above unity. This result implies that gas traced by [\ion{C}{1}] \((2\mathrm{-}1)\) and CO \((7\mathrm{-}6)\) in the inner part of the torus is warm and dense. By comparison, [\ion{C}{1}] \((1\mathrm{-}0)\) has been detected in a nuclear spiral structure and a rotating ring with a radius of 2 pc (CND) of our Galaxy, thereby showing evidence of abundant atomic carbon in the vicinity of the SMBH \citep{TNK26}. However, the density and temperature of gas probed by [\ion{C}{1}] are higher at the nucleus of NGC 613, possibly because of higher concentration of molecular gas and its excitation due to AGN feedback.

The measured CO\((3\mathrm{-}2)/(2\mathrm{-}1)\) line intensity ratio in the broad component that traces the extended outflow is \(W_\mathrm{CO(3-2)}/W_\mathrm{CO(2-1)}\approx1\), which is similar to the total ratio in the CND. This supports the scenario that the molecular wind is warm and launched from the torus region.

\subsection{NGC 1808}

\begin{figure}[ht!]
\epsscale{1.1}
\plotone{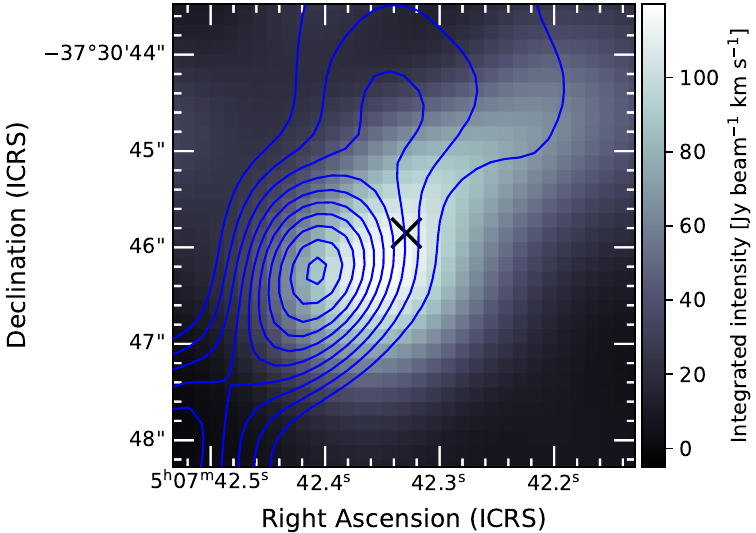}
\caption{Integrated intensity map of the blue-shifted component (\(v_\mathrm{LSR}<v_\mathrm{sys}-120~\mathrm{km~s}^{-1}\)) in CO \((2\mathrm{-}1)\) in NGC 1808 with contours at \((0.1,0.2,0.3,0.4,0.5,0.6,0.7,0.8,0.9,0.98)\times\mathcal{I}_\mathrm{CO(2-1)}^\mathrm{max}\). The grey scale shows CO \((2\mathrm{-}1)\) intensity integrated over all velocities. The nucleus is indicated by an \textsf{x}-symbol.\label{fig:wing4}}
\end{figure}

The spectra of [\ion{C}{1}] \((2\mathrm{-}1)\) and CO \((7\mathrm{-}6)\) at the nucleus of NGC 1808 are more complex and do not show unambiguous evidence of high-velocity components (Figure \ref{fig:spec}). However, the spectrum in Figure \ref{fig:radex2} (top right) clearly shows that CO \((2\mathrm{-}1)\) and \((3\mathrm{-}2)\) lines exhibit wings at velocities that are blue-shifted with respect to the systemic velocity by more than \(200~\mathrm{km~s}^{-1}\). This feature was noticed in previous studies and interpreted as a molecular outflow from the central region \citep{Sal16,Sal17}. Figure \ref{fig:wing4} shows that the blue-shifted component seen in CO \((2\mathrm{-}1)\) has a peak in a region that is \(\approx50~\mathrm{pc}\) southeast from the nucleus and is extended. The absence of evidence of outflows from the nucleus at the scale of \(10~\mathrm{pc}\) suggests that the winds are probably driven by starburst feedback and launched from a wider region.

\section{Summary}\label{sec:sum}

We have presented the first ALMA Band 10 observations of [\ion{C}{1}] \(({^3\mathrm{P}_2}\mathrm{-}{^3\mathrm{P}_1})\) and CO \((J=7\mathrm{-}6)\) toward the central regions of the Seyfert galaxies NGC 613 and NGC 1808 at unprecedented resolution of \(10~\mathrm{pc}\). The main results are summarized below.

\begin{enumerate}

\item{The [\ion{C}{1}] \((2\mathrm{-}1)\) and CO \((7\mathrm{-}6)\) lines, and 800-GHz continuum were detected toward the CND of both targets and exhibit maximum values at the positions of the galaxy nuclei that coincide with the molecular tori. While the two lines show similar distributions, CO \((7\mathrm{-}6)\) is more compact and is concentrated in the torus. The [\ion{C}{1}]\((2\mathrm{-}1)\)/CO\((7\mathrm{-}6)\) integrated-flux ratio is found to be \(\approx0.3\) at the nucleus and \(\approx0.5\mathrm{-}0.6\) over the entire CND.}

\item{The peak of the [\ion{C}{1}] integrated-intensity ratio is \(W_\mathrm{[CI](2-1)}/W_\mathrm{[CI](1-0)}\approx1.1\) (\(\mathrm{K~km~s}^{-1}\) scale) in the CNDs of both galaxies. Under the approximation of LTE and optically thin emission, this implies an excitation temperature of \(\approx60~\mathrm{K}\) and a column density of \(N_\mathrm{C}\approx(7\mathrm{-}8)\times10^{18}~\mathrm{cm}^{-2}\) within the central 39 pc.}

\item{To further analyze the gas physical conditions, we used the non-LTE radiative transfer code RADEX. The intensity ratio of the two [\ion{C}{1}] lines constrains the density and kinetic temperature of molecular gas in both nuclei to \(n_\mathrm{H_2}>10^3~\mathrm{cm}^{-3}\) and \(T_\mathrm{k}>50~\mathrm{K}\), respectively. To probe the C/CO abundance ratio, we also simultaneously modeled the intensities of [\ion{C}{1}] and multiple CO lines in the central part of the CNDs. The resulting abundance ratio is \(\approx1.6\) in NGC 613 and \(\approx0.63\) in NGC 1808 at the resolution of 63 pc. The solutions are obtained for dense (\(n_\mathrm{H_2}\sim10^{3\mathrm{-}4}\mathrm{cm}^{-3}\)) and warm (\(T_\mathrm{k}\sim80\mathrm{-}250~\mathrm{K}\)) molecular gas. In the central 34 pc of NGC 613, C/CO and \(n_\mathrm{H_2}\) are estimated to increase to \(\approx2.0\) and \(\sim10^{4.4}~\mathrm{cm}^{-3}\). C/CO is found to increase with \(T_\mathrm{k}\) and decrease with \(n_\mathrm{H_2}\) in the CND of NGC 1808, indicating that the abundance of C is enhanced relative to CO in regions that are less shielded from ambient radiation field.}

\item{We also investigated the possible mechanisms that produce the observed line intensities and C/CO abundance. The measured [\ion{C}{1}]\((2\mathrm{-}1)/(1\mathrm{-}0)\) line ratios can be reproduced by numerical models of PDRs with a FUV radiation field of \(G_0\sim10^{2\mathrm{-}4}\) and CR ionization rate of \(\zeta\sim10^{-16}~\mathrm{s}^{-1}\), and by models of XDRs with a moderate/low hard X-ray flux of \(F_\mathrm{X}\sim0.1\mathrm{-}1~\mathrm{erg~s^{-1}~cm^{-2}}\). The FUV radiation field is higher in NGC 613 compared to NGC 1808 in all PDR models (plane-parallel and clumpy). A combined analysis of [\ion{C}{1}] and CO line ratios is in favor of the PDR scenario.}

\item{The spectra extracted from the CND of NGC 613 could be fitted with two Gaussian components. The main component, that traces the bulk of the molecular torus, is clearly visible in all lines, whereas broad components are identified in CO \((2\mathrm{-}1)\), \((3\mathrm{-}2)\), and \((7\mathrm{-}6)\) lines. The high velocities seen in CO \((2\mathrm{-}1)\) and \((3\mathrm{-}2)\) are consistent with an AGN-driven wind blowing from the CND that has been reported in previous works. On the other hand, the secondary components in CO \((7\mathrm{-}6)\) and [\ion{C}{1}] \((2\mathrm{-}1)\) likely originate in the inner part of the torus. Assuming circular motion, we estimated the mass of the supermassive black hole at the center of NGC 613 to be \(\sim2\times10^7~M_\sun\). The line intensity ratios are \(W_\mathrm{[CI](2-1)}/W_\mathrm{[CI](1-0)}>1\) and \(W_\mathrm{CO(3-2)}/W_\mathrm{CO(2-1)}\approx1\) in the broad components, indicating high excitation of the gas in the torus and outflow.}

\item{The spectra of [\ion{C}{1}] \((2\mathrm{-}1)\) and CO \((7\mathrm{-}6)\) toward the nucleus of NGC 1808 do not reveal unambiguous evidence of outflows in the torus region. The molecular outflow in NGC 1808, reported earlier based on large-scale CO \((1\mathrm{-}0)\), \((2\mathrm{-}1)\), and \((3\mathrm{-}2)\) observations at lower resolution is probably starburst-driven and launched from a wider region.}

\end{enumerate}

\newpage

\begin{acknowledgments}
The authors thank the referee for many constructive comments that helped us improve the manuscript. This paper makes use of the following ALMA data: ADS/JAO.ALMA \(\#\)2013.1.00911.S, ADS/JAO.ALMA \(\#\)2015.1.01487.S, ADS/JAO.ALMA \(\#\)2015.1.00404.S, ADS/JAO.ALMA \(\#\)2017.1.00236.S, ADS/JAO.ALMA \(\#\)2017.1.00984.S, ADS/JAO.ALMA \(\#\)2017.1.01671.S, and ADS/JAO.ALMA \(\#\)2023.1.01177.S. ALMA is a partnership of ESO (representing its member states), NSF (USA), and NINS (Japan), together with NRC (Canada), MOST and ASIAA (Taiwan), and KASI (Republic of Korea), in cooperation with the Republic of Chile. The Joint ALMA Observatory is operated by ESO, AUI/NRAO, and NAOJ. This research has made use of the NASA/IPAC Extragalactic Database (NED), which is operated by the Jet Propulsion Laboratory, California Institute of Technology, under contract with the National Aeronautics and Space Administration. Based on observations made with the NASA/ESA Hubble Space Telescope, and obtained from the Hubble Legacy Archive, which is a collaboration between the Space Telescope Science Institute (STScI/NASA), the Space Telescope European Coordinating Facility (ST-ECF/ESA) and the Canadian Astronomy Data Centre (CADC/NRC/CSA). The {\it HST} data used in this paper can be found in MAST: \dataset[10.17909/w87f-wz42]{http://dx.doi.org/10.17909/w87f-wz42}. Dragan Salak was supported by the ALMA Japan Research Grant of NAOJ ALMA Project, NAOJ-ALMA-402.
\end{acknowledgments}
\software{CASA \citep{TCT22}, Matplotlib \citep{Hun07}, PDRT \citep{PW23}, RADEX \citep{vdT07}, SciPy \citep{Vir20}}

\end{document}